\documentclass[pdflatex,sn-mathphys-num]{sn-jnl}% Math and Physical Sciences Numbered Reference Style
\usepackage{graphicx}%
\usepackage{multirow}%
\usepackage{amsmath,amssymb,amsfonts,mathtools}%
\usepackage{amsthm}%
\usepackage[title]{appendix}%
\usepackage{xcolor}%
\usepackage{textcomp}%
\usepackage{manyfoot}%
\usepackage{booktabs}%
\usepackage{algorithm}%
\usepackage{algorithmicx}%
\usepackage{algpseudocode}%
\usepackage{listings}%
\usepackage{subcaption}
\usepackage{xr}
\theoremstyle{thmstyleone}%
\theoremstyle{thmstyletwo}%

\theoremstyle{thmstylethree}%

\begin{document}

\title[Article Title]{Stochastic Thermodynamics of Score Matching in Diffusion Models}


\author[1]{\fnm{Xuehao} \sur{Ding}}\email{xding@flatironinstitute.org}

\author[2,3,4]{\fnm{H. T.} \sur{Quan}}\email{htquan@pku.edu.cn}

\author[1]{\fnm{Yuhai} \sur{Tu}}\email{ytu@flatironinstitute.org}

\affil[1]{\orgdiv{Flatiron Institute}, \orgname{Simons Foundation}}

\affil[2]{\orgdiv{School of Physics}, \orgname{Peking University}}
\affil[3]{\orgname{Collaborative Innovation Center of Quantum Matter, Peking University}}
\affil[4]{\orgname{Frontiers Science Center for Nano-optoelectronics, Peking University}}

%TC:ignore
%\abstract{The score-based diffusion model is a prominent generative-AI framework that can sample from complex probability distributions in very high-dimensional spaces. The forward diffusion process evolves the data sample into Gaussian noise, and the reverse sampling process is trained to reverse the probability flow to turn the noise into generated samples. In this work, we study the diffusion model and the score-matching objective function under the framework of stochastic thermodynamics. We introduce time-asymmetry entropy production (TAEP), defined by the dynamics of the forward and reverse processes of the diffusion model, and show that it satisfies the fluctuation theorems. Remarkably, Hyvärinen’s implicit score-matching kernel naturally emerges in the fluctuating TAEP. The average TAEP is exactly proportional to the score-matching objective, and its variance characterizes the unevenness of sampling. As direct consequences of the fluctuation theorem, we quantitatively explain why diffusion models can generate samples with high diversity, and how stochastic gradient descent biases the optimization of the score-matching objective toward more generalizable solutions. This work reveals the entropic nature of the score-matching objective for diffusion models, thereby elucidating fundamental statistical-mechanical principles underlying generative AI.}

\abstract{Score-based diffusion models are a powerful class of generative AI systems capable of sampling from complex, high-dimensional probability distributions. Their dynamics consist of a forward diffusion process that transforms data into noise and a learned reverse process that reconstructs data by reversing the probability flow. Here, we develop a stochastic thermodynamic framework for diffusion models and their score-matching objective. We introduce a trajectory-dependent quantity, time-asymmetry entropy production (TAEP), defined from the forward and reverse diffusion dynamics, and show that it obeys exact fluctuation theorems. Remarkably, Hyvärinen’s implicit score-matching kernel emerges naturally as a fluctuating component of TAEP, while the average TAEP is exactly proportional to the score-matching objective. We further show that fluctuations of TAEP quantify sampling unevenness and provide a thermodynamic measure of data-manifold coverage. These results yield a quantitative explanation for the superior sampling diversity of diffusion models and reveal a thermodynamic mechanism by which stochastic gradient descent favors flatter, more generalizable solutions. By uncovering the entropic nature of score matching, our work establishes fundamental statistical-mechanical principles underlying diffusion-based generative AI.}

\maketitle

\section{Main}

The diffusion model~\cite{sohl2015deep} is a state-of-the-art generative machine learning framework with broad applications in computer vision, audio generation, natural language processing, and more. Mathematically, the diffusion model provides a general approach capable of learning the structure of sophisticated data distributions in high-dimensional spaces, e.g., the distribution of naturalistic images. Intuitively, the forward process continuously adds noise to the data sample until it becomes pure Gaussian noise. The learned reverse sampling process is expected to act as a denoiser at each time step to turn the noise back into a generated data sample. That is, the reverse process should reverse the arrow of time to evolve the Gaussian distribution into the target data distribution.

The original work on the diffusion model~\cite{sohl2015deep} was inspired by the Jarzynski equality~\cite{jarzynski1997nonequilibrium} and the fluctuation theorem~\cite{seifert2005entropy}, which conceptualize forward and backward trajectories of thermodynamic systems. Despite the fact that the vanilla diffusion model was formulated as a variational inference problem, subsequent efforts~\cite{song2019generative,ho2020denoising, songscore} demonstrated that the reverse sampling process could be understood as a reverse-time stochastic differential equation (SDE)~\cite{anderson1982reverse} and the objective function for training diffusion models could be reformulated via score matching~\cite{hyvarinen2005estimation}.

%The SDE perspective makes it feasible to rethink the diffusion model as a physical system and study its thermodynamic relationships during the learning and inference phases under the framework of non-equilibrium statistical physics. Recently, there has been some work done in this direction. In particular, Ref.~\cite{yu2025nonequilbrium} applies Seifert's fluctuation theorem for total entropy production (EP)~\cite{seifert2005entropy} to the diffusion model using an analytically tractable model. Ref.~\cite{ikeda2025speed} derives the speed-accuracy trade-off relationship for the diffusion model and studies the optimal learning protocol. Refs.~\cite{biroli2023generative,raya2023spontaneous,biroli2024dynamical,ambrogioni2025statistical} study spontaneous symmetry breaking under the reverse dynamics of the diffusion model. However, none of these works directly relates thermodynamic concepts to the score-matching objective, which is the canonical objective function for training diffusion models and of central interest to the AI community.

The SDE perspective allows diffusion models to be interpreted as stochastic physical systems and analyzed within the framework of nonequilibrium statistical physics. Recent studies have explored this direction from several angles, including entropy production~\cite{yu2025nonequilbrium}, speed--accuracy trade-offs and optimal learning protocols~\cite{ikeda2025speed}, and spontaneous symmetry breaking in reverse diffusion dynamics~\cite{biroli2023generative,raya2023spontaneous,biroli2024dynamical,ambrogioni2025statistical}. However, none of these works establishes a direct link between thermodynamic principles and the score-matching objective, the canonical objective for training diffusion models and a cornerstone of modern generative AI.

In this work, we develop a stochastic thermodynamic framework for diffusion models. Motivated by trajectory-based formulations of work~\cite{jarzynski1997nonequilibrium, sekimoto2010stochastic} and entropy production (EP)~\cite{seifert2005entropy}, we introduce time-asymmetry entropy production (TAEP), which we will show is a trajectory-based score-matching objective. After reviewing key concepts from stochastic thermodynamics, in particular path-integral representations of EP and fluctuation theorems, we formulate the forward and reverse diffusion processes using Langevin equations and derive explicit expressions for both fluctuating and average TAEP. We establish the fluctuation theorems for TAEP and demonstrate its fundamental connection to score matching. In the special case where the neural network represents an exact score field, TAEP directly quantifies the discrepancy between the generated and target distributions.

Our framework leads to two main consequences. First, the fluctuation theorem provides a thermodynamic explanation for the high sampling diversity and broad data-manifold coverage of diffusion models, linking these properties to fluctuations of TAEP. Second, a corollary of the fluctuation theorem predicts an architecture-agnostic positive relationship between SGD noise strength and loss-landscape curvature, suggesting that stochastic optimization naturally drives the score-matching objective toward flatter, and hence more generalizable, minima. Numerical experiments support these theoretical predictions.

\section{Brief Review of Stochastic Thermodynamics}

In this section, we briefly review key results in stochastic thermodynamics relevant to our work~\cite{seifert2012stochastic, esposito2010three_a, van2015ensemble}. Consider a small system in contact with a heat reservoir. The dynamics of the system are stochastic due to thermal fluctuations. In the continuous setting, the evolution of the probability distribution of the system state can be described by the Fokker-Planck equation:
\begin{equation}
\frac{\partial p(x,t)}{\partial t}=-\nabla\cdot[\mu F(x)p(x,t)-\mu k_B T\nabla p(x,t)],
\label{fokker planck}
\end{equation}
where $\mu$ is the mobility, $F(x)$ is the force, $k_B$ is the Boltzmann constant, and $T$ is the temperature of the heat reservoir. %For simplicity, here we neglect the external control parameter, which is often considered in the literature. Besides, it is also common to describe the system dynamics using discrete master equations, and all results in this section hold for both the continuous and discrete settings.

Consider a fixed time interval $t\in [0,\tau]$, over which the probability distribution of the system evolves from $p(x,0)$ to $p(x,\tau)$. The total entropy production of each trajectory $x([t])$ satisfies the following relation~\cite{seifert2005entropy, maes2003time}.
\begin{equation}
\label{total entropy production}
\Delta s_{tot}[x([t])]=k_B\log \frac{p[x([t])]}{\bar{p}[\bar{x}([t])]},
\end{equation}
where $p$ is the forward probability density measure of trajectories with the initial distribution $p(x, 0)$, $\bar{p}$ is the backward probability density measure under the same dynamics as the forward process but with the initial distribution $p(x,\tau)$, and $\bar{x}([t])$ denotes the time-reversed trajectory of $x([t])$, i.e., $x(t)=\bar{x}(\tau-t)$. The total EP can be decomposed into the EP of the system $\Delta s_{sys}$ and the EP of the heat bath $\Delta s_{bath}$, which satisfy
\begin{equation}
    \label{system and heat bath}
    \Delta s_{tot}=\Delta s_{sys}+\Delta s_{bath},
\end{equation}
\begin{equation}
\label{system entropy}
\begin{aligned}
\Delta s_{sys}[x([t])]=(-k_B\log p[x(\tau),\tau]) - (-k_B\log p[x(0),0]),
\end{aligned}
\end{equation}
\begin{equation}
\Delta s_{bath}[x([t])]=k_B\log \frac{p[x([t])|x(0)]}{\bar{p}[\bar{x}([t])|x(\tau)]}.
\end{equation}
The ensemble averages of these trajectory-level quantities agree with the standard macroscopic EPs:
\begin{eqnarray}
\langle \Delta s_{sys}\rangle=\Delta S_{sys},\; \langle \Delta s_{bath}\rangle=\Delta S_{bath},\; \langle \Delta s_{tot}\rangle=\Delta S_{tot}.
\end{eqnarray}

From equation \eqref{total entropy production}, one can derive the following integral fluctuation theorem and detailed fluctuation theorem for total EP via path integration~\cite{van2015ensemble}.
\begin{eqnarray}
\label{ift}
&\langle\exp(-\frac{\Delta s_{tot}}{k_B})\rangle=1,\\
\label{dft}
& \frac{p(\Delta s_{tot})}{\bar{p}(-\Delta s_{tot})}=\exp(\frac{\Delta s_{tot}}{k_B}).
\end{eqnarray}
Here, the probability measures of EP are defined as
\begin{eqnarray}
p(\sigma)=\int \mathcal{D}x([t]) \delta\left(\sigma-k_B\log \frac{p[x([t])]}{\bar{p}[\bar{x}([t])]}\right)p[x([t])],\\
\bar{p}(\sigma)=\int \mathcal{D}\bar{x}([t]) \delta\left(\sigma-k_B\log \frac{\bar{p}[\bar{x}([t])]}{p[x([t])]}\right)\bar{p}[\bar{x}([t])].\\
\end{eqnarray}
Applying Jensen's inequality to equation \eqref{ift} yields the second law of thermodynamics
\begin{equation}
\label{2nd law}
    \Delta S_{tot}=\langle \Delta s_{tot}\rangle \geq 0.
\end{equation}
The detailed fluctuation theorem, equation \eqref{dft}, states that an entropy increase is exponentially more probable than an entropy decrease.

Aside from splitting the total EP into system EP and reservoir EP (equation \eqref{system and heat bath}), an alternative way is to split the total EP into non-adiabatic EP and adiabatic EP, closely related to the concepts of excess heat and housekeeping heat~\cite{oono1998steady, hatano2001steady, esposito2010three_a,esposito2010three_b, van2010three}. Intuitively, the non-adiabatic EP is the extra EP relative to the corresponding adiabatic process. Thus, the non-adiabatic EP of an adiabatic process is zero. The non-adiabatic EP of each trajectory $x([t])$ satisfies the following relation.
\begin{equation}
\label{na entropy production}
\Delta s_{na}[x([t])]=k_B\log \frac{p[x([t])]}{\bar{p}^\dagger[\bar{x}([t])]},
\end{equation}
where $\bar{p}^\dagger$ denotes the probability density measure of the backward trajectory under the dual dynamics with the initial condition $p(x,\tau)$. The force of the dual dynamics is given by
\begin{eqnarray}
\label{dual force}
F^\dagger(x)=-F(x)+2k_BT\nabla \log p^{st}(x),
\end{eqnarray}
where $p^{st}(x)$ denotes the stationary distribution of the forward dynamics, and the diffusion term of the dual dynamics is the same as that of the forward dynamics. Similarly to total EP, non-adiabatic EP also satisfies an integral fluctuation theorem and a detailed fluctuation theorem~\cite{esposito2010three_a}:
\begin{eqnarray}
\label{ift_na}
&\langle\exp(-\frac{\Delta s_{na}}{k_B})\rangle=1,\\
\label{dft_na}
& \frac{p(\Delta s_{na})}{\bar{p}^{\dagger}(-\Delta s_{na})}=\exp(\frac{\Delta s_{na}}{k_B}),
\end{eqnarray}
where the dual-reversed probability measure of EP is defined as
\begin{equation}
    \bar{p}^\dagger(\sigma)=\int \mathcal{D}\bar{x}([t]) \delta\left(\sigma-k_B\log \frac{\bar{p}^\dagger[\bar{x}([t])]}{p[x([t])]}\right)\bar{p}^\dagger[\bar{x}([t])].
\end{equation}
With a change of variables, equation \eqref{ift_na} becomes the well-known Hatano-Sasa relation~\cite{hatano2001steady}. 

\section{Time-asymmetry entropy production in diffusion models}
\subsection{Diffusion model described by Langevin equations}

\begin{figure}[ht]
  \centering
  \includegraphics[width=0.75\textwidth]{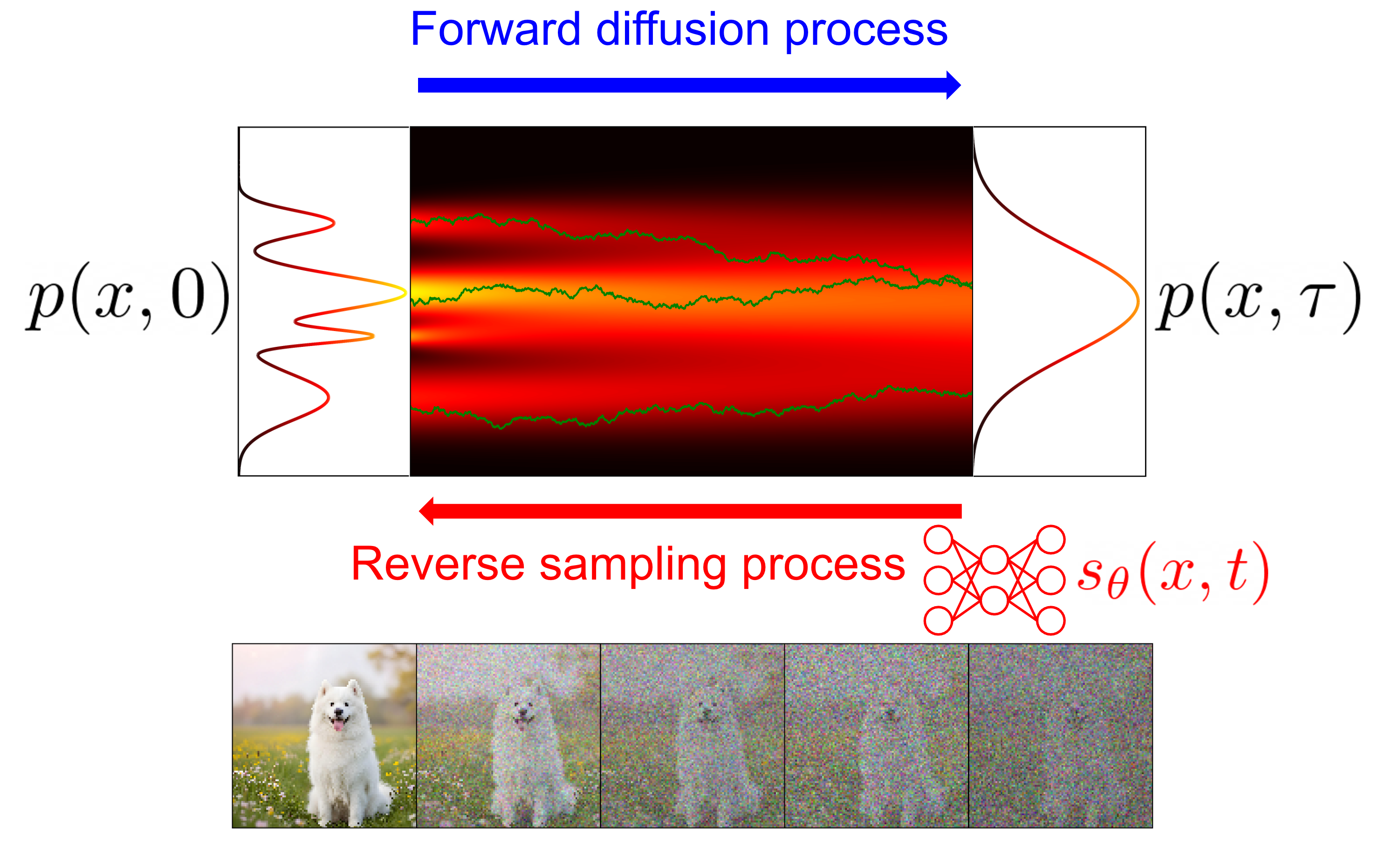}
  \caption{Schematic of the diffusion model.}
  \label{schematic}
\end{figure}

A diffusion model consists of a forward diffusion process and a reverse sampling process (Fig. \ref{schematic}). The forward process transforms a generically intractable data distribution into a tractable Gaussian distribution by adding noise to data samples. The reverse process has to be trained to evolve the Gaussian distribution back into the target data distribution by denoising corrupted samples.

We model the forward process of the diffusion model using a high-dimensional overdamped Langevin equation.
\begin{equation}
\label{forward process}
    dx=\mu F(x)dt+\sqrt{2\mu k_BT} \circ dW_t,
\end{equation}
where $t\in [0,\tau]$, $x\in \mathbb{R}^n$ denotes the corrupted data sample in the forward diffusion process, $F(x)\in \mathbb{R}^n$ denotes the drift force, and $W_t\in \mathbb{R}^n$ is a Wiener process. Following the convention of stochastic thermodynamics, here we adopt Stratonovich's stochastic calculus~\cite{sekimoto2010stochastic}. In the literature on diffusion models, the setting $F(x)=0$ is termed variance-exploding, in which the variance of the distribution grows without bound. In contrast, $F(x)=-\frac{k_BT}{\sigma^2}x$ is termed variance-preserving with the stationary variance equal to $\sigma^2$, which corresponds to an Ornstein–Uhlenbeck process. 

The evolution of the probability distribution $p(x,t)$ is governed by the Fokker-Planck equation, equation \eqref{fokker planck}, and $p(x,0)$ represents the target data distribution. The probability current of the forward process is given by
\begin{equation}
    J(x,t)=p(x,t)[\mu F(x)-\mu k_BT\nabla \log p(x,t)].
\end{equation}

The dynamics of the reverse sampling process can also be described by an overdamped Langevin equation, albeit non-autonomous~\cite{songscore}:
\begin{equation}
\label{reverse process}
    d\widetilde{x}=\mu [-F(\widetilde{x})+2k_BTs_\theta(\widetilde{x},\widetilde{t})]d\widetilde{t}+\sqrt{2\mu k_BT} \circ d\widetilde{W}_t,
\end{equation}
where $\widetilde{t}\in[0,\tau]$, $\widetilde{x}\in \mathbb{R}^n$ denotes the intermediate denoised sample in the reverse process, $s_\theta(\widetilde{x},\widetilde{t})\in \mathbb{R}^n$ is the denoising neural network with parameters $\theta$, $\widetilde{W}_t\in \mathbb{R}^n$ is the Wiener process in the reverse process. The only difference from the dual dynamics (equation \ref{dual force}) is that the stationary score is replaced by the denoising network. Let $\widetilde{p}_\theta(\widetilde{x}, \widetilde{t})$ denote the parameter-dependent probability distribution in the reverse process. Since the purpose of the reverse process is to transform the final distribution of the forward process into the data distribution, we naturally have the following.
\begin{equation}
\widetilde{p}_{\theta}(x, 0)=p(x,\tau).
\end{equation}
The probability current of the backward process is given by
\begin{equation}
\begin{aligned}
\widetilde{J}_\theta(\widetilde{x}, \widetilde{t})=\widetilde{p}_\theta(\widetilde{x}, \widetilde{t})[-\mu F(\widetilde{x})+2\mu k_BTs_\theta(\widetilde{x},\widetilde{t})-\mu k_BT\nabla \log \widetilde{p}_\theta(\widetilde{x}, \widetilde{t})].
\end{aligned}
\end{equation}
When there exists a set of parameters $\theta^*$ such that
\begin{equation}
    s_{\theta^*}(x,t)=\nabla \log p(x,\tau-t),
\end{equation}
equation \eqref{reverse process} becomes the reverse-time SDE proposed by Ref.~\cite{anderson1982reverse}, which reverses the probability current of the forward process. That is, 
\begin{equation}
\label{reverse the probability flow}
    \widetilde{J}_{\theta^*}(x, t)=-J(x,\tau-t),\quad \widetilde{p}_{\theta^*}(x, t)=p(x,\tau-t).
\end{equation}
Thus, assuming that the expressivity of the network architecture is sufficient, the fully trained network $s_\theta(x,t)$ is expected to converge to the score $\nabla \log p(x,\tau-t)$ and $\theta^*$ is the optimal set of parameters.

In the literature on diffusion models, the canonical objective function to minimize is the score-matching objective defined as the mean squared error of the estimated score integrated over time~\cite{songscore}, which is 
\begin{equation}
\label{score matching objective}
    \mathcal{L}_{SM}(\theta)\coloneq \int_0^\tau\langle||s_\theta(x,\tau-t)-\nabla \log p(x,t)||^2\rangle_{x\sim p(x,t)} dt.
\end{equation}
Clearly, the optimum satisfies $\mathcal{L}_{SM}(\theta^*)=0$. In the following, we reveal the entropic nature of the score-matching objective.

%\subsection{Entropic nature of the score-matching objective}

\subsection{Entropic nature of score matching and fluctuation theorems}

%We now introduce the trajectory-dependent TAEP for the diffusion model as follows, similar to equations \eqref{total entropy production}\eqref{na entropy production}.
%\begin{equation}
%\label{nr entropy production}
%\Delta s_{ta}[x([t])]=k_B\log \frac{p[x([t])]}{\widetilde{p}_\theta[\bar{x}([t])]},
%\end{equation}
%where the dynamics of the backward process are given by the reverse sampling process, equation \eqref{reverse process}, of the diffusion model. Note that equation \eqref{nr entropy production} is not the conventional Radon-Nikodym derivative, but rather a ratio between trajectory densities in the Onsager-Machlup sense as commonly adopted in stochastic thermodynamics~\cite{lau2007state}. For a fully (perfectly) trained network, it follows from equation \eqref{reverse the probability flow} that
%\begin{equation}
%    p[x([t])]=\widetilde{p}_{\theta^*}[\bar{x}([t])],
%\end{equation}
%which means that the TAEP vanishes for every trajectory. In practice, the training is not perfect, and TAEP measures the discrepancy between the trajectory densities in the forward evolution and the trained reverse process.

We now introduce the trajectory-dependent time-asymmetry entropy production (TAEP) for diffusion models, analogous to the total and non-adiabatic entropy productions satisfying equations~\eqref{total entropy production} and \eqref{na entropy production}, respectively:
\begin{equation}
\label{nr entropy production}
\Delta s_{ta}[x([t])]=k_B\log \frac{p[x([t])]}{\widetilde{p}_\theta[\bar{x}([t])]},
\end{equation}
where $\widetilde{p}_\theta$ denotes the trajectory density measure under the reverse sampling dynamics, equation~\eqref{reverse process}. We emphasize that equation~\eqref{nr entropy production} is not a conventional Radon--Nikodym derivative, but rather the ratio between trajectory densities in the Onsager--Machlup sense, as commonly used in stochastic thermodynamics~\cite{lau2007state}.

For a perfectly trained model, equation \eqref{reverse the probability flow} implies
\begin{equation}
p[x([t])]=\widetilde{p}_{\theta^*}[\bar{x}([t])],
\end{equation}
so that the TAEP vanishes identically for every trajectory. In practice, however, the learned score field is only approximate. The TAEP therefore quantifies the time asymmetry between the trajectory densities in the forward evolution and the learned reverse dynamics, providing a trajectory-level measure of model error.

Next, we explicitly evaluate equation \eqref{nr entropy production}. Using the method developed in stochastic thermodynamics~\cite{chernyak2006path, lau2007state}, we find (see the Supplement, Section I):
\begin{equation}
\label{TAEP expression}
\begin{aligned}
&\Delta s_{ta}[x([t])]=\mu k_B^2T\int_{0}^\tau [l(\theta;x(t),t)-l(\theta^*;x(t),t)]dt\\
&+k_B\sqrt{2\mu k_B T}\int_{0}^{\tau} [s_\theta(x(t),\tau-t)-s_{\theta^*}(x(t),\tau-t)]^\top dW_t,
\end{aligned}
\end{equation}
\begin{equation}
\label{score matching kernel}
    l(\theta;x,t)\coloneq ||s_\theta(x,\tau-t)||^2+ 2\nabla \cdot s_\theta(x,\tau-t).
\end{equation}
Importantly, $l(\theta;x,t)$ is precisely Hyvärinen’s implicit score-matching kernel introduced in the original work on score matching~\cite{hyvarinen2005estimation}. The last term consists of the projection of the score estimation error onto the corrupting noise~\cite{song2020sliced}.

%It follows from equation \eqref{nr entropy production} that the fluctuating TAEP satisfies the fluctuation theorems, which is one of main theoretical results in our paper:
%\begin{eqnarray}
%\label{ift_nr}
%&\langle\exp(-\frac{\Delta s_{ta}}{k_B})\rangle=1,\\
%\label{dft_nr}
%& \frac{p(\Delta s_{ta})}{\widetilde{p}_\theta(-\Delta s_{ta})}=\exp(\frac{\Delta s_{ta}}{k_B}),
%\end{eqnarray}
%where the backward probability measure of EP is defined as
%\begin{equation}
%    \widetilde{p}_\theta(\sigma)=\int \mathcal{D}\bar{x}([t]) \delta\left(\sigma-k_B\log \frac{\widetilde{p}_\theta[\bar{x}([t])]}{p[x([t])]}\right)\widetilde{p}_\theta[\bar{x}([t])].
%\end{equation}
%Numerically, we verify the integral fluctuation theorem in Fig. \ref{toy model}, and we find that for diffusion models trained on natural images, the family of $\Delta s_{ta}$ distributions during optimization closely resembles the family of work/entropy distributions of a thermodynamic system approaching quasi-staticity (Fig. \ref{distribution} and Fig. 8 in Ref.~\cite{jarzynski1997equilibrium}).

Employing the path-integral technique in stochastic thermodynamics~\cite{seifert2005entropy, esposito2010three_a, maes2003time}, one can show from equation~\eqref{nr entropy production} that the fluctuating TAEP obeys the integral and detailed fluctuation theorems: 
\begin{eqnarray}
\label{ift_nr}
&\langle\exp(-\frac{\Delta s_{ta}}{k_B})\rangle=1,\\
\label{dft_nr}
& \frac{p(\Delta s_{ta})}{\widetilde{p}_\theta(-\Delta s_{ta})}=\exp(\frac{\Delta s_{ta}}{k_B}),
\end{eqnarray}
%\begin{eqnarray} \label{ift_nr} &\left\langle e^{-\Delta %s_{ta}/k_B}\right\rangle = 1,\ \label{dft_nr} &\frac{p(\Delta %s_{ta})}{\widetilde{p}\theta(-\Delta s{ta})} = e^{\Delta %s_{ta}/k_B}, \end{eqnarray} 
which constitute one of the central theoretical results of this work. Here, the backward probability measure of EP is defined as
\begin{equation}
    \widetilde{p}_\theta(\sigma)=\int \mathcal{D}\bar{x}([t]) \delta\left(\sigma-k_B\log \frac{\widetilde{p}_\theta[\bar{x}([t])]}{p[x([t])]}\right)\widetilde{p}_\theta[\bar{x}([t])].
\end{equation}

We numerically verify the integral fluctuation theorem in Fig.~\ref{toy model}. Moreover, for diffusion models trained on natural-image datasets, we find that the evolution of the TAEP distribution during training closely mirrors the evolution of work or EP distributions in thermodynamic systems approaching the quasi-static limit (Fig.~\ref{distribution}; see also Fig.~8 of Ref.~\cite{jarzynski1997equilibrium}). This observation suggests a deep analogy between diffusion-model optimization and nonequilibrium thermodynamic relaxation.

%We further evaluate the ensemble-averaged TAEP. Remarkably, we find that the average TAEP is exactly proportional to the score-matching objective (see the Supplement, Section II).
%\begin{equation}
%\label{EP equivalent to loss}
%\Delta S_{ta}=\langle\Delta s_{ta}\rangle= \mu k_B^2T\mathcal{L}_{SM}(\theta).
%\end{equation}
%Equivalently, the TAEP rate is given by
%\begin{equation}
%\frac{dS_{ta}}{dt}=\mu k_B^2T\langle||s_\theta(x,\tau-t)-\nabla \log p(x,t)||^2\rangle_{x\sim p(x,t)},
%\end{equation}
%which is the score-matching objective at time $t$. These results indicate that the score-matching objective shares a mathematical structure very similar to that of entropy production in stochastic thermodynamics, and $\Delta s_{ta}$ can serve as a trajectory score-matching objective. We emphasize that the trajectory-based score-matching objective is a novel concept we introduce, and we will show in Section 4 that its fluctuations play a significant role in the context of generative AI.

We next evaluate the ensemble-averaged TAEP. Remarkably, the average TAEP is exactly proportional to the score-matching objective (see the Supplement, Section II):
\begin{equation}
\label{EP equivalent to loss}
\Delta S_{ta}=\langle\Delta s_{ta}\rangle= \mu k_B^2T\mathcal{L}_{SM}(\theta).
\end{equation}
Moreover, the TAEP rate is given by
\begin{equation}
\label{EP equivalent to loss rate}
\frac{dS_{ta}}{dt}=\mu k_B^2T\langle||s_\theta(x,\tau-t)-\nabla \log p(x,t)||^2\rangle_{x\sim p(x,t)},
\end{equation}
which coincides with the instantaneous score-matching objective.

As explicitly demonstrated in equations~\eqref{TAEP expression},~\eqref{EP equivalent to loss}, and~\eqref{EP equivalent to loss rate}, TAEP provides a novel trajectory-based formulation of the score-matching objective in diffusion models. As we show in Section~4, fluctuations of the trajectory TAEP play a crucial role in understanding the behavior and performance of diffusion-based generative models.

%\subsection{Transfer learning, generalization, and near-optimality}

\subsection{TAEP measures target-to-generated distribution mismatch}

To understand the physical meaning of TAEP, we consider the case in which the neural network $s_\theta(x,\tau -t)$ represents an exact score field:
\begin{equation}
\label{exact score field}
    s_\theta(x,\tau-t)=\nabla \log q(x,t),
\end{equation}
where $q(x,t)$ is a time-dependent density governed by the dynamics of the same forward process as $p(x,t)$. It follows that $q(x,0)$ is the distribution of generated samples. Even though equation \eqref{exact score field} is not generally true, it holds as a valid approximation in several scenarios including transfer learning, generalization, and when the trained model is near-optimal (see the Supplement, Section III). %{\color{red} Maybe it is better to move the following itemized list to SI or get rid of them entirely as we don't really deal with some of the things mentioned such as transfer learning...} :
%\begin{enumerate}
%\item Transfer learning, which is a machine learning paradigm that leverages knowledge acquired from a pre-trained source domain to enhance the learning efficiency and predictive performance of a model on a related but distinct task. Suppose that there is a dataset $\mathcal{P}$ with distribution $p(x,0)$ and a related dataset $\mathcal{Q}$ with distribution $q(x,0)$. Suppose that a diffusion model has been fully trained on $\mathcal{Q}$, i.e., equation \eqref{exact score field} is satisfied. Now we want to train another diffusion model to learn the distribution of $\mathcal{P}$. Instead of training from scratch, we start from the existing network to transfer knowledge about $\mathcal{Q}$ to the new diffusion model.
%\item Generalization. Similarly to transfer learning, the model has been fully trained on the train set $\mathcal{Q}$, and now we test the model on the test set $\mathcal{P}$.
%\item Near-optimality. When $\theta$ is in the neighborhood of the optimum $\theta^*$, $s_\theta(x,\tau-t)$ can be well approximated as its projection onto the score-field space~\cite{lai2023fp}, which we denote by $\nabla \log q(x,t)$.
%\end{enumerate}

In this case, we find (see the Supplement, Section III) that the trajectory TAEP is given by
\begin{equation}
    \Delta s_{ta}[x([t])]=k_B\int_0^\tau \partial_\nu \log\frac{q[x(t),t]}{p[x(t),t]}\circ dx^\nu,
\end{equation}
where we adopt Einstein notation and covariant indices: $\partial_0\coloneq\partial_t$, $x^0\coloneq t$. It follows that the ensemble-averaged TAEP rate can be expressed as the familiar form of a sum over fluxes times forces:
\begin{equation}
\label{nr EP rate for transfer learning}
\frac{d S_{ta}}{dt}=k_B\int \mathcal{J}^\nu(x,t)N_\nu(x,t)dx,
\end{equation}
\begin{equation}
\label{thermodynamic force}
N_\nu=\partial_\nu\log\frac{q(x,t)}{p(x,t)},
\end{equation}
where $\mathcal{J}\coloneq (p,J)$ denotes the spacetime probability current, and $N_\nu$ denotes the corresponding thermodynamic force. Mathematically, equations \eqref{nr EP rate for transfer learning} and \eqref{thermodynamic force} reduce to the non-adiabatic EP rate when $q(x,t)$ is the stationary distribution of the forward dynamics~\cite{van2010three}.

%Using integration by parts and assuming that the boundary term vanishes, equation \eqref{nr EP rate for transfer learning} can be alternatively written as the derivative of a KL divergence:
%\begin{equation}
%    \frac{d S_{ta}}{dt}=-k_B\frac{d}{dt}D_{KL}[p(x,t)||q(x,t)],
%\end{equation}
%which is known to be non-negative~\cite{csiszar1967information}. Intuitively, $p(x,t)$ and $q(x,t)$ both converge to the stationary distribution in the forward process, and thus their KL divergence decreases.

Importantly, in the limit $\tau\rightarrow \infty$, the TAEP depends only on the initial condition:
\begin{equation}
\label{infinite tau EP}
    \Delta s_{ta}[x([t])]=k_B\log \frac{p[x(0),0]}{q[x(0),0]},\quad \Delta S_{ta}=k_B D_{KL}[p(x,0)||q(x,0)],
\end{equation}
which directly measures the discrepancy between the target and generated distributions.

\section{Applications}  

In this section, we apply the integral fluctuation theorem to study the performance of diffusion models in terms of sampling diversity and learning dynamics. Our study reveals how the variance of the TAEP distribution characterizes sampling unevenness in diffusion models and how it is implicitly minimized by optimization. Furthermore, we show that in diffusion models, the SGD noise covariance is positively correlated with the Hessian of the objective function, thereby driving the optimization toward flatter minima. %can bias the system towards flatter more generalizable solutions. %      We further elucidate the role of its second-order moment and the relationship between the first two moments. Additionally, the SGD noise–Hessian alignment is justified as a corollary.

\subsection{Var\texorpdfstring{$(\Delta s_{ta})$}{Delta sta} as a signature of diversity in generative sampling}

In heat engines, average work and average entropy production define efficiency. However, assessing reliability requires knowing how much these fluctuating quantities vary~\cite{ding2018measurement,pietzonka2018universal}. 
We can apply this same logic to diffusion models. The standard score-matching objective only measures average sample quality. In contrast, higher-order moments of the TAEP reveal sampling unevenness across the data distribution.

\begin{figure}[ht]
  \centering
  \includegraphics[width=1.\textwidth]{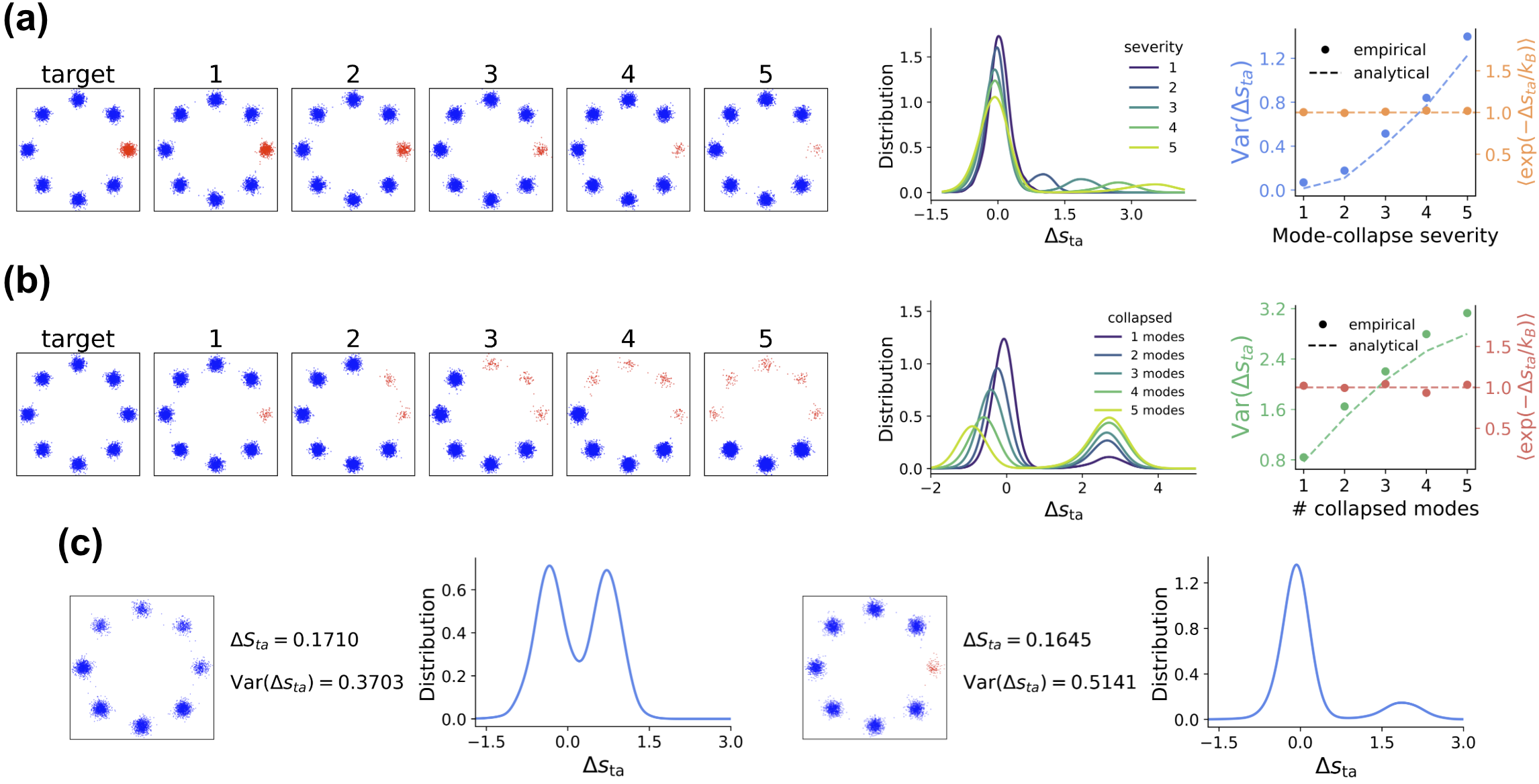}
  \caption{Diffusion model on 2D Gaussian-mixture data. (a) Left: The first panel shows the target data distribution, eight Gaussians on a 2D ring with equal weights. The other panels show the distributions generated by five diffusion models with increasing collapse severity of a single mode (colored in red). Middle: Density of $\Delta s_{ta}$ for each mode-collapse severity. Right: Variance of TAEP increasing with the severity of mode collapse and numerical verification of the integral fluctuation theorem. Analytical values from equations \eqref{relative entropy variance} and \eqref{ift_nr} are shown. (b) Similar to (a) but for increasing numbers of collapsed modes. (c) The distributions generated by two diffusion models with similar average TAEPs. The one with severe mode collapse has significantly higher variance. The corresponding TAEP densities are shown.}
  \label{toy model}
\end{figure}

A prime example of uneven sampling is mode collapse~\cite{salimans2016improved}, in which a model concentrates its probability on just a few data types and fails to generate diverse samples, leaving other valid data modes underrepresented.
To show how the TAEP distribution detects and quantifies mode collapse, we tested diffusion models on 2D Gaussian mixtures. We constructed models with different levels of mode collapse by fitting them to biased datasets (see Methods). As shown in Fig. \ref{toy model}a, the TAEP distributions clearly reflect the severity of mode collapse (middle panel). Specifically, the variance, Var$(\Delta s_{ta})$, increases as mode collapse gets worse (right panel). Remarkably, every single model still strictly obeys the fluctuation theorem (right panel). The same trend of increasing Var$(\Delta s_{ta})$ with the severity of mode collapse is observed for increasing numbers of collapsed modes, as shown in Fig. \ref{toy model}b. While the mean TAEP, $\Delta S_{ta}$, measures the average mismatch between the target and generated distributions, the variance provides further information on sampling heterogeneity (Fig. \ref{toy model}c).

We note that the above numerical results become analytically clear when considering the special case in which the neural network represents a score field (equation \eqref{exact score field}) and $\tau\rightarrow \infty$, as assumed in Section 3.3. It follows from equation \eqref{infinite tau EP} that the variance of the TAEP is given by the relative entropy variance:
\begin{equation}
\label{relative entropy variance}
    \text{Var}(\Delta s_{ta})=k_B^2\text{Var}_{p(x,0)}\left[\log\frac{p(x,0)}{q(x,0)}\right],
\end{equation}
a concept used in quantum information~\cite{li2014second}. 

The relative entropy variance complements KL divergence by tracking how unevenly errors are spread across samples, rather than just measuring the mean error level. Thus, $\text{Var}(\Delta s_{ta})$ can be used to distinguish between two diffusion models that have the exact same score-matching objective, i.e., even if their average first-order errors match, their second-order structures can be completely different (Fig. \ref{toy model}c). For example, one model might be consistently mediocre across the entire data space. The other model might fit some regions perfectly while completely missing others, which is what happens during mode collapse or other forms of nonuniform coverage. Interestingly, this exact trade-off between sample quality and diversity is a major focus of research on classifier-free diffusion guidance~\cite{ho2021classifier}.

%Relative entropy variance complements KL divergence by measuring how unevenly the discrepancy between the target and generated distributions is distributed across samples, beyond its mean level. Consequently, the variance of the TAEP can distinguish two diffusion models that exhibit similar values of the score-matching objective, the first-order discrepancy, while differing substantially in second-order structure: one may be consistently mediocre over the whole data manifold, whereas the other may fit some regions well while severely underrepresenting others, as in mode collapse or other forms of nonuniform coverage (Fig. \ref{toy model}c). Interestingly, a similar trade-off between quality and diversity has received considerable attention in the context of classifier-free diffusion guidance~\cite{ho2021classifier}.

We note that the first two moments of the TAEP can be related through the cumulant expansion of the integral fluctuation theorem, equation \eqref{ift_nr}:
\begin{equation}
\label{fdt}
    \Delta S_{ta}=\frac{1}{2k_B}\text{Var}(\Delta s_{ta})+\text{higher-order cumulants},
\end{equation}
which corresponds to the fluctuation-dissipation theorem for total EP when the higher-order cumulants are negligible, valid near equilibrium~\cite{jarzynski1997nonequilibrium}. Equation \eqref{fdt} indicates that the mean TAEP is proportional to the variance up to higher-order terms. Despite the higher-order correction, we provide a rigorous mathematical statement in terms of the asymptotic mean-variance relation as the mean approaches zero (see the Supplement, Section IV). This relation suggests that optimizations that reduce the score-matching objective also implicitly reduce the variance. 

\begin{figure}[ht]
  \centering
  \begin{subfigure}[t]{0.36\linewidth}
    \caption{}
    \includegraphics[width=\columnwidth]{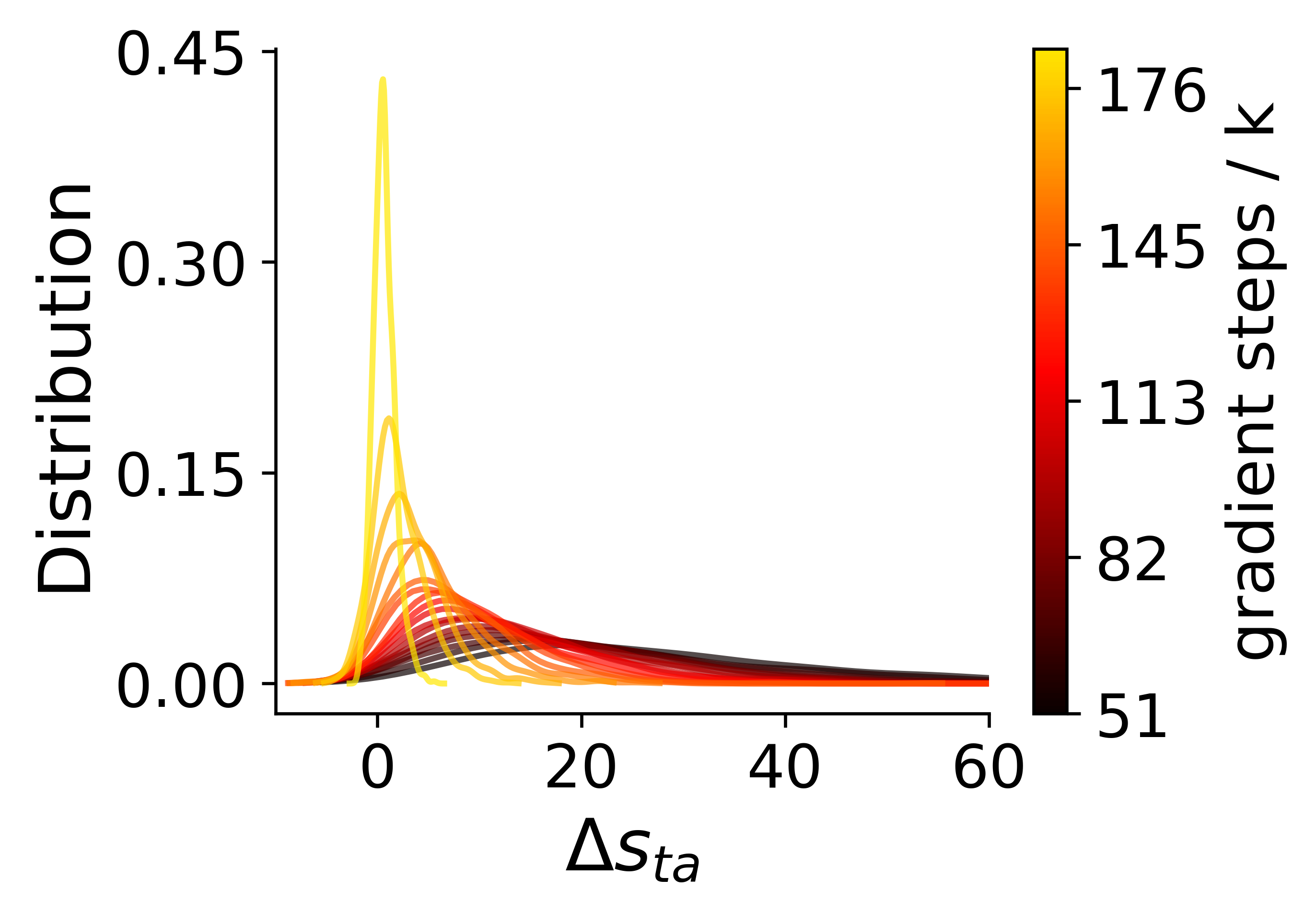}
    \label{distribution}
    \end{subfigure}
    \begin{subfigure}[t]{0.31\linewidth}
    \caption{}
    \includegraphics[width=\columnwidth]{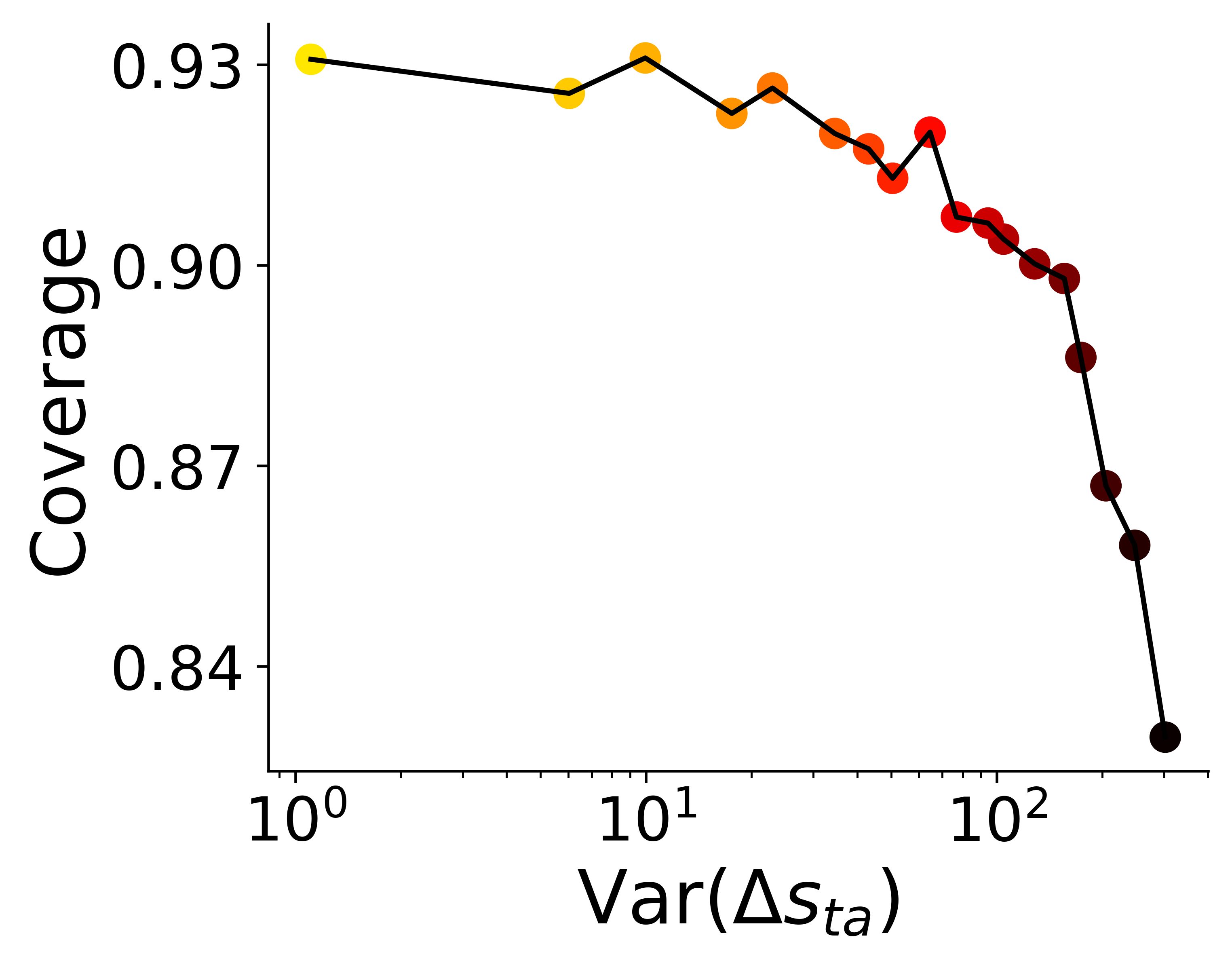}
    \label{coverage}
    \end{subfigure}
  \begin{subfigure}[t]{0.31\linewidth}
    \caption{}
    \includegraphics[width=\columnwidth]{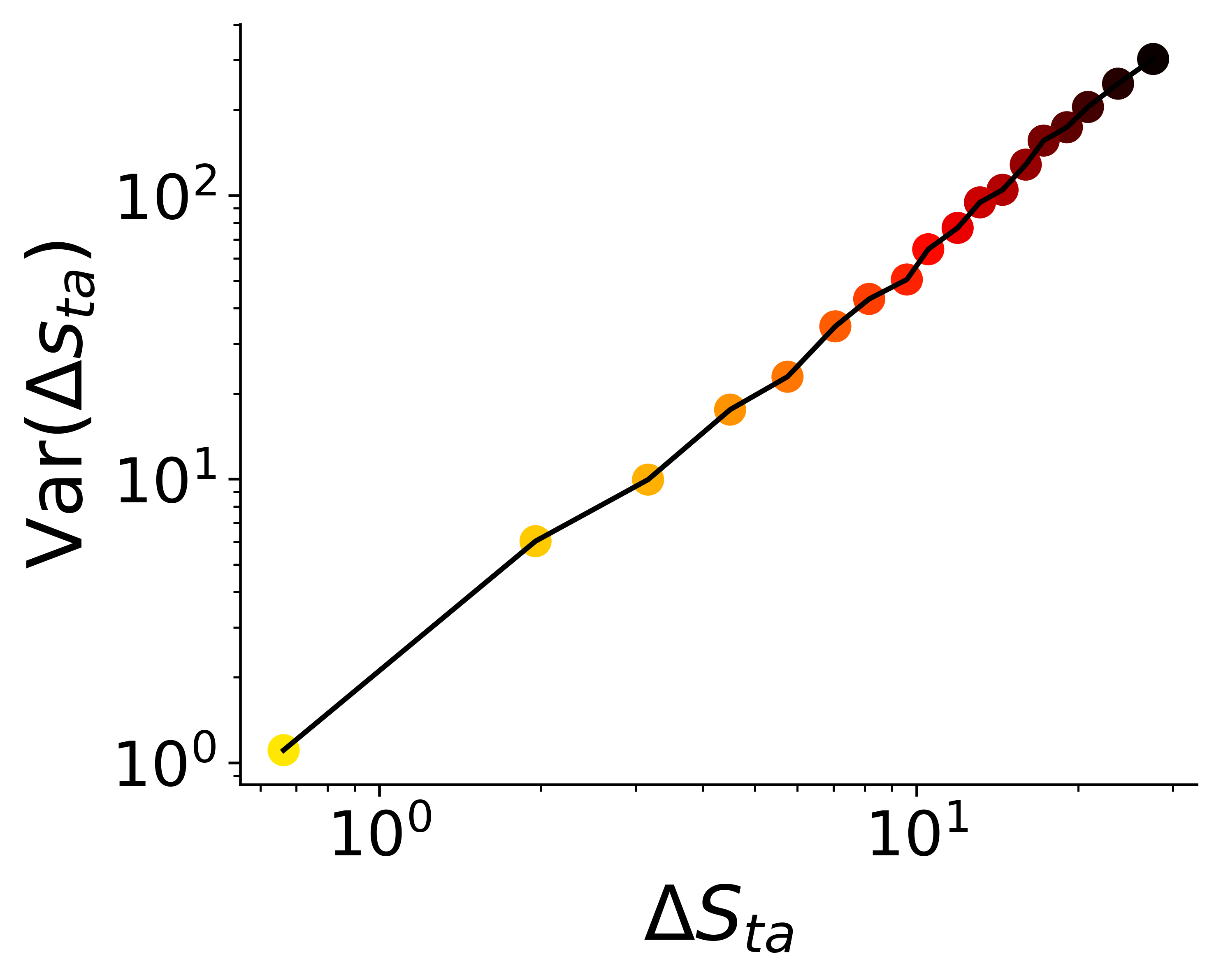}
    \label{var_mean}
    \end{subfigure}
  \caption{Diffusion model on CIFAR-10. (a) The distribution of TAEP along the training process. (b) The coverage of the data manifold by generated samples increases as the variance of TAEP is reduced during the optimization process. (c) The variance of TAEP decreases along with the mean during the optimization process.}
  \label{var steps}
\end{figure}

We numerically test our analytical results for diffusion models trained on natural images. The variance of the TAEP distribution tracks the coverage of the data manifold by generated samples, as defined in Ref.~\cite{naeem2020reliable} (Fig. \ref{coverage}). We further verify that the variance indeed decreases along with the mean during training (Fig. \ref{var_mean}), as detailed in Methods. Under the interpretation of the variance, the relationship between the first two moments of $\Delta s_{ta}$ naturally explains why diffusion models exhibit substantially milder mode collapse and better coverage of the full data distribution compared to earlier generative models, including GANs, VAEs, and autoregressive models~\cite{nichol2021improved, dhariwal2021diffusion}.

\subsection{Landscape-dependent SGD Noise in Diffusion Models}

For neural networks trained with stochastic gradient descent, the SGD noise is defined as the difference between the gradient of the per-sample or per-batch loss w.r.t. parameters and that of the mean loss. It has been found in multi-layer perceptron models that the covariance of SGD noise is positively correlated with the Hessian of the objective function w.r.t. model parameters, which characterizes the curvature of the loss landscape~\cite{jastrzebski2018three}. Specifically, in sharper directions of the loss landscape with larger Hessian eigenvalues, the SGD noise strength is also stronger. This loss-landscape-dependent anisotropic SGD noise is critical in biasing the optimization process toward finding flatter minima~\cite{feng2021inverse, yang2023stochastic}, which have been shown to have better generalization performance~\cite{feng2023activity}. However, the positive correlation between the SGD noise covariance and the Hessian of the objective function has so far been shown only in simple model architectures~\cite{zhu2019anisotropic, li2020hessian, zhang2026superlinearrelationshipsgdnoise}.

For diffusion models and the score-matching objective, we note that the positive correlation between Hessian and SGD noise covariance is a direct consequence of the fluctuation theorem, agnostic of the model architecture. By taking the second-order gradient of equation \eqref{fdt} w.r.t. $\theta$, we obtain
\begin{equation}
\label{IVF}
\begin{aligned}
    \mathbf{H}_\theta (\Delta S_{ta})=&\frac{1}{2k_B}\text{Cov}(\nabla_\theta \Delta s_{ta}, \nabla_\theta \Delta s_{ta})\\
    &+\frac{1}{2k_B}\text{Cov}(\Delta s_{ta}, \mathbf{H}_\theta( \Delta s_{ta}))+\text{higher-order terms},
\end{aligned}
\end{equation}
where $\mathbf{H}_\theta$ denotes the Hessian w.r.t. model parameters. According to equation \eqref{EP equivalent to loss}, the LHS of the above equation is proportional to the Hessian of the score-matching objective. Since $\Delta s_{ta}$ is a per-trajectory loss, the first term of the RHS is the covariance of the SGD noise. The same statement also holds for the sample-wise covariance, as we show that the per-sample loss is a coarse-grained mean of the per-trajectory loss, so their variances exhibit a positive linear relationship (see the Supplement, Section V and Fig. S1). 

\begin{figure}[ht]
  \centering
  \begin{subfigure}[t]{0.3\linewidth}
    \caption{}
    \includegraphics[width=\columnwidth]{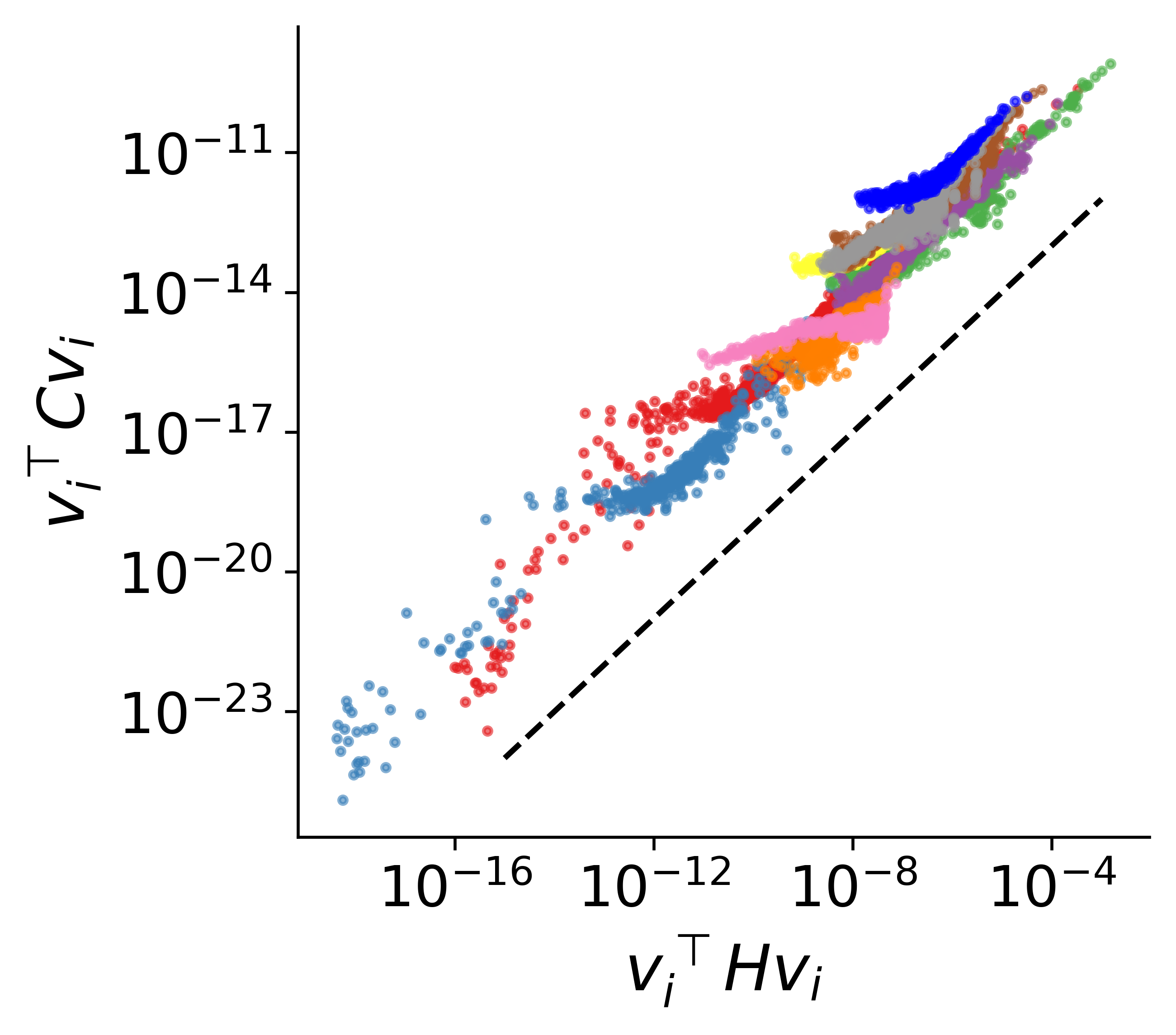}
    \label{ivf layer}
    \end{subfigure}
    \begin{subfigure}[t]{0.33\linewidth}
    \caption{}
    \includegraphics[width=\columnwidth]{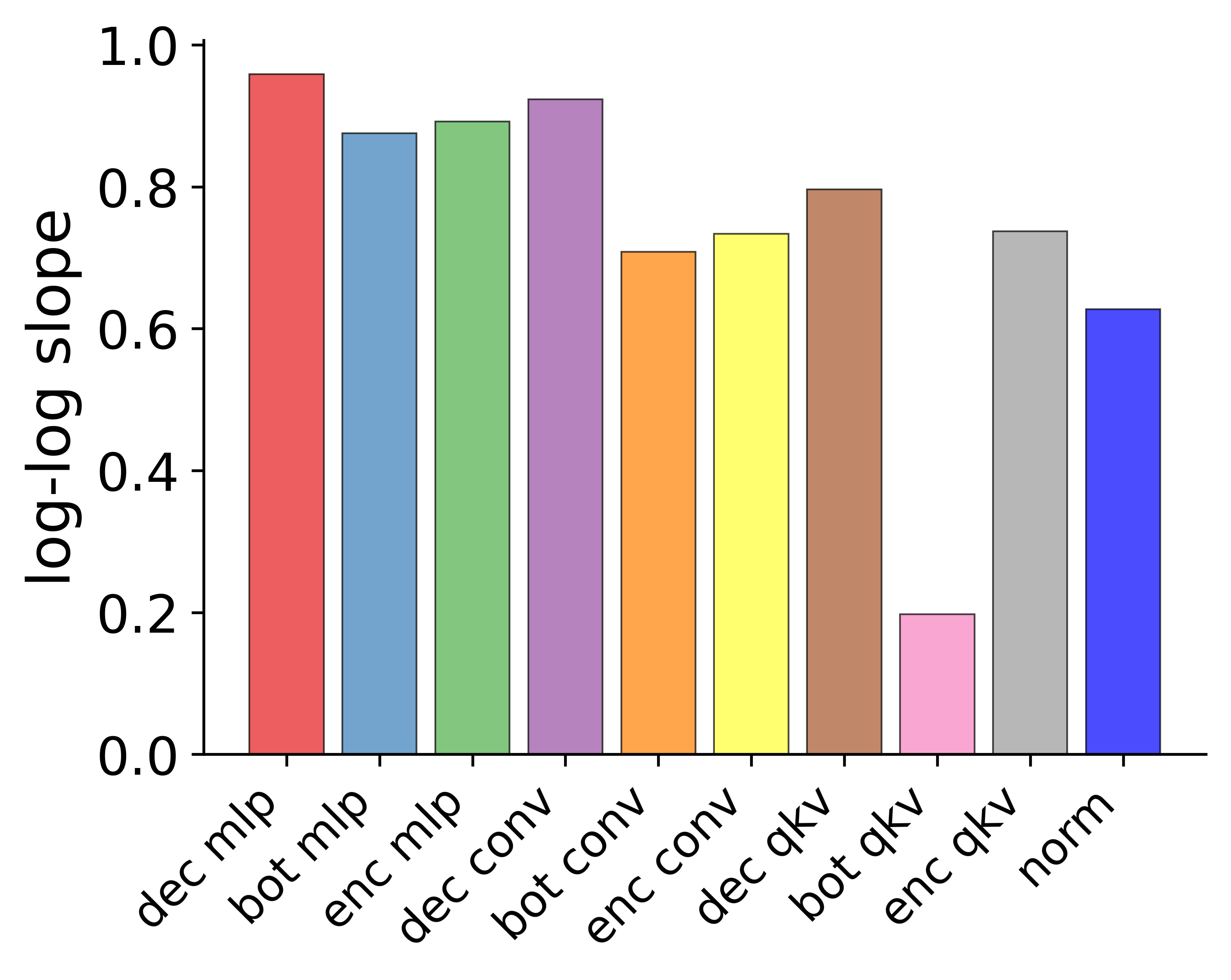}
    \label{ivf slope}
    \end{subfigure}
  \begin{subfigure}[t]{0.35\linewidth}
    \caption{}
    \includegraphics[width=\columnwidth]{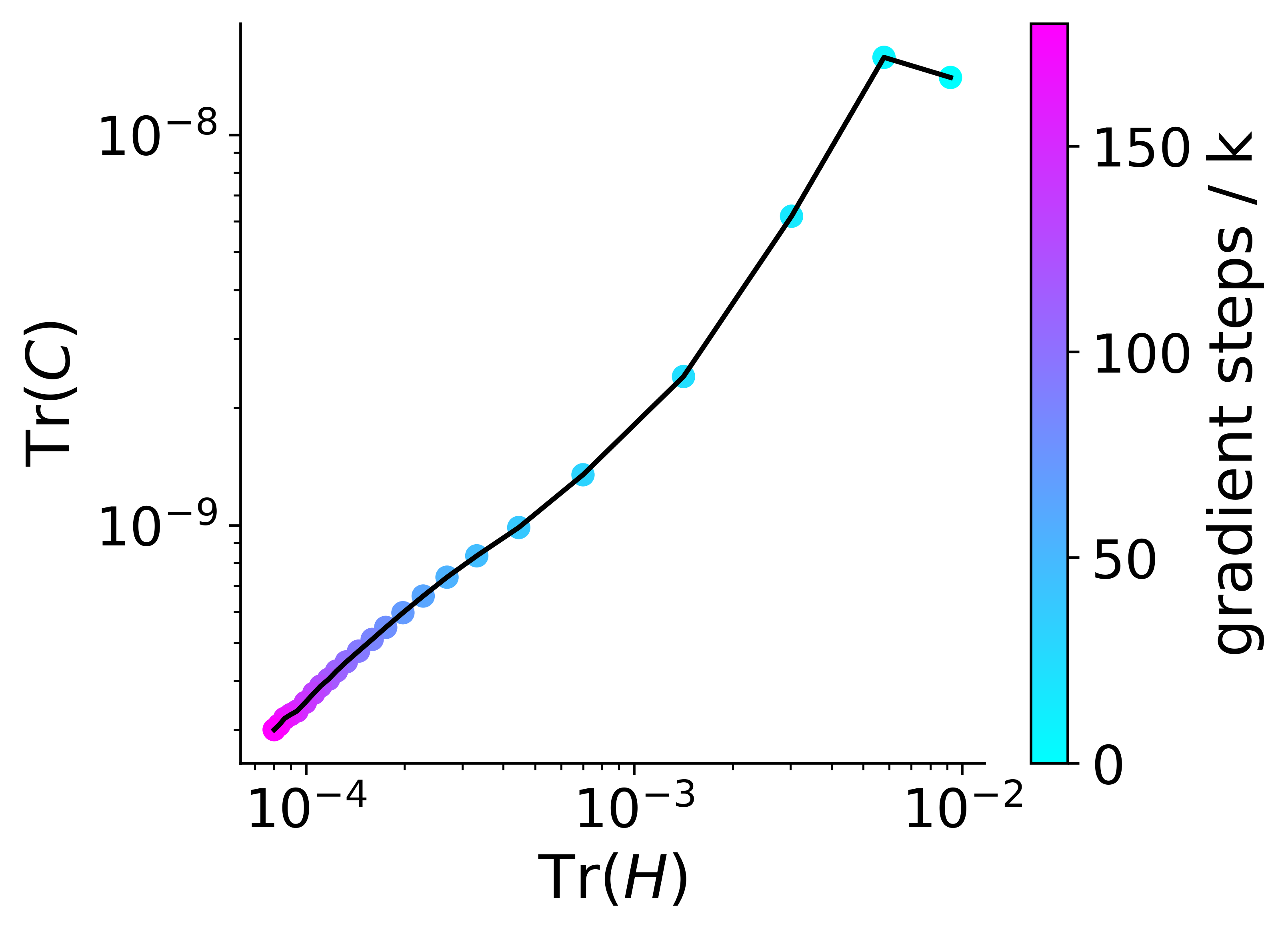}
    \label{ivf trace}
    \end{subfigure}
  \caption{Numerical results on the relationship between Hessian (loss curvature) and SGD noise covariance for diffusion model on CIFAR-10. We denote the Hessian matrix by $H$, the normalized top eigenvectors of $H$ by $v_i$, and the covariance matrix of SGD noise by $C$. (a) Eigenvalues of the Hessian versus Rayleigh quotients of the SGD noise covariance at the eigenvectors of the Hessian. Results for parameters in the MLP/convolutional/attention/normalization layers in the decoder/bottleneck/encoder of the U-Net are plotted, with the color coding given in (b). A dashed line with log-log slope equal to $1$ is shown for comparison. (b) Log-log slope of the SGD noise covariance $v_i^\top C v_i$ versus the Hessian eigenvalue $v_i^\top H v_i$, fitted by linear regression for each layer type. (c) Trace of the Hessian versus that of the SGD noise covariance for models at various gradient steps.}
  \label{ivf}
\end{figure}

We test the theoretical results using diffusion models trained on CIFAR-10 (see Methods). As shown in Fig. \ref{ivf layer}, the SGD noise covariance is positively correlated with the Hessian for parameters in various layers. The general trend across different layers of the network follows a power law with an exponent $\sim 1$, although the exponent within each layer is smaller (Fig. \ref{ivf slope}). We also evaluate the overall strength of the SGD noise and the sharpness of the loss landscape characterized by the traces of the noise covariance and the Hessian, respectively, along the training trajectory. As shown in Fig. \ref{ivf trace}, the landscape-dependent SGD noise and sharpness of the loss landscape decrease with training. Our results demonstrate both theoretically and empirically that the SGD noise covariance is positively correlated with the Hessian in diffusion models, which can systematically bias the optimization of the score-matching objective toward flatter, more robust minima with better generalization.

\section{Outlook}

The diffusion model has two-level dynamics: 1. the dynamics of the forward and reverse processes; 2. the learning dynamics of the model during optimization. Thus, a natural next step is to discover the physical laws that govern the coupling between these two sets of dynamics. Since we find that the score-matching objective can be interpreted as entropy production, one may formulate the learning process under the general principle of minimal entropy production.

This work shows that the mathematical structure of the score-matching objective is very similar to that of non-adiabatic EP in classical stochastic thermodynamics, which has been studied for decades. Recent findings show that the concept of non-adiabatic entropy can be generalized to quantum physics~\cite{esposito2009nonequilibrium, horowitz2014equivalent,manzano2018quantum}. Therefore, our work also opens the possibility of constructing novel objective functions for generative AI models inspired by non-classical physics.

Stochastic thermodynamics provides a powerful framework for understanding fluctuations, irreversibility, and nonequilibrium processes. Despite substantial efforts, however, its real-world applications remain relatively limited. Our work offers a concrete example of how the concepts and principles of stochastic thermodynamics can be applied to practical scenarios. 

In general, this work builds a conceptual bridge that allows statistical physicists to bring their expertise and rich toolbox to the field of generative AI. The results presented here suggest that this interface is a promising direction for future research, and we anticipate that further developments along these lines will yield increasingly significant discoveries.

%TC:ignore
\section*{Methods}
\setcounter{section}{0}
\setcounter{subsection}{0}

\section{Diffusion model on Gaussian mixture}

\subsection{Gaussian-mixture data distribution}

We used a two-dimensional Gaussian-mixture model to test the interpretation of TAEP fluctuations as a diagnostic of sampling unevenness. The target distribution was an equally weighted mixture of $K=8$ isotropic Gaussian components,
\begin{equation}
    p(x,0)
    =
    \frac{1}{K}
    \sum_{k=1}^{K}
    \mathcal{N}(x;m_k,\sigma_0^2 I_2),
    \qquad x\in\mathbb{R}^2 .
\end{equation}
The component centers were placed uniformly on a circle,
\begin{equation}
    m_k
    =
    R
    \left(
    \cos\frac{2\pi(k-1)}{K},
    \sin\frac{2\pi(k-1)}{K}
    \right),
    \qquad k=1,\ldots,K .
\end{equation}
We used $R=4$ and $\sigma_0=0.25$. Samples were drawn directly from the Gaussian mixture without additional preprocessing or data augmentation.

\subsection{Diffusion protocol and score-network architecture}

We trained variance-preserving diffusion models following the standard DDPM procedure~\cite{ho2020denoising}. The noise schedule was linear in the variance-preserving SDE parameter, with $\beta_{\min}=0.1$, $\beta_{\max}=20$, and training times sampled uniformly from $[10^{-3},1]$. The time-dependent noise schedule can be absorbed into a monotone reparameterization of diffusion time to match the Langevin equation in the main text.

The score network was a multilayer perceptron $s_\theta(x,t):\mathbb{R}^2\times[0,1]\rightarrow\mathbb{R}^2$. The input consisted of the noisy sample $x_t$ concatenated with a sinusoidal time embedding of $t$. The network had three hidden layers of width $128$, SiLU nonlinearities, and a two-dimensional output. The same architecture was used for all controlled mode-collapse settings.

\subsection{Optimization}

The networks were trained with Adam using a learning rate of $10^{-3}$ and a batch size of $2048$. We clipped the global gradient norm at $1.0$. Unless stated otherwise, each model was trained for $5\times10^4$ optimization steps. An exponential moving average (EMA) of the parameters was maintained during training and used for sampling and EP evaluation.

\subsection{Controlled construction of mode-collapse models}

To obtain controlled mode-collapse models, we trained diffusion models on biased Gaussian mixtures while evaluating all EP statistics relative to the uniform target distribution $p_0$. For a model with $m$ underrepresented modes, the training distribution was
\begin{equation}
    q^{(m,\epsilon)}(x,0)
    =
    \sum_{k=1}^{K}
    w_k^{(m,\epsilon)}
    \mathcal{N}(x;m_k,\sigma_0^2 I_2).
\end{equation}
The first $m$ modes were assigned probability $\epsilon$ each,
\begin{equation}
    w_k^{(m,\epsilon)}=\epsilon,
    \qquad k=1,\ldots,m,
\end{equation}
and the remaining $K-m$ modes shared the rest of the probability mass equally,
\begin{equation}
    w_k^{(m,\epsilon)}
    =
    \frac{1-m\epsilon}{K-m},
    \qquad k=m+1,\ldots,K.
\end{equation}
For the single-mode-collapse experiment, we used $m=1$ and varied $\epsilon=0.09,0.05,0.02,0.01,0.005$. For the multi-mode-collapse experiment, we fixed $\epsilon=0.01$ and varied $m=1,\ldots,5$.

\subsection{Trajectory-level TAEP evaluation}

The TAEP of each trajectory was numerically evaluated following equation \eqref{TAEP expression}. The optimal score of the target distribution is available analytically. Under the variance-preserving forward process, the marginal distribution at time $t$ is still a Gaussian mixture,
\begin{equation}
    p(x,t)
    =
    \frac{1}{K}
    \sum_{k=1}^{K}
    \mathcal{N}
    \left(
    x;
    \sqrt{\bar{\alpha}(t)}\,m_k,
    v_t I_2
    \right),
\end{equation}
where
\begin{equation}
    v_t
    =
    \bar{\alpha}(t)\sigma_0^2
    +
    1-\bar{\alpha}(t).
\end{equation}
Here $\bar{\alpha}(t)$ is the cumulative signal factor of the variance-preserving diffusion process. Define
\begin{equation}
    \mu_k(t)=\sqrt{\bar{\alpha}(t)}\,m_k
\end{equation}
and the posterior responsibility
\begin{equation}
    r_k(x,t)
    =
    \frac{
    \exp\left[-\|x-\mu_k(t)\|^2/(2v_t)\right]
    }{
    \sum_{j=1}^{K}
    \exp\left[-\|x-\mu_j(t)\|^2/(2v_t)\right]
    } .
\end{equation}
The exact optimal score is then
\begin{equation}
    s_{\theta^*}(x,t)
    =
    \nabla\log p(x,t)
    =
    \sum_{k=1}^{K}
    r_k(x,t)
    \left[
    -\frac{x-\mu_k(t)}{v_t}
    \right].
\end{equation}
We used this analytical score as the optimal model in the trajectory-level EP calculation. Since the system is two-dimensional, the divergence term was evaluated exactly by automatic differentiation.

\subsection{Analytical relative entropy variance}

Equation \eqref{relative entropy variance} in the main text shows that in the exact-score-field and long-time limit,
\begin{equation*}
    {\rm Var}\left(\frac{\Delta s_{\rm ta}}{k_B}\right)
    =
    {\rm Var}_{p(x,0)}
    \left[
    \log\frac{p(x,0)}{q(x,0)}
    \right].
\end{equation*}
For well-separated Gaussian components with identical covariance and different mode weights, this reduces to a discrete mode-weight expression. If the target weights are uniform, $p_k=1/K$, and the generated weights are $w_k$, then
\begin{equation}
    {\rm Var}\left(\frac{\Delta s_{\rm ta}}{k_B}\right)
    \simeq
    \frac{1}{K}
    \sum_{k=1}^{K}
    \left(
    \log\frac{1/K}{w_k}
    -
    \frac{1}{K}
    \sum_{j=1}^{K}
    \log\frac{1/K}{w_j}
    \right)^2 .
\end{equation}
For the controlled $m$-mode-collapse construction,
\begin{equation}
    w_1=\cdots=w_m=\epsilon,
    \qquad
    w_{m+1}=\cdots=w_K=\frac{1-m\epsilon}{K-m},
\end{equation}
which gives
\begin{equation}
    {\rm Var}\left(\frac{\Delta s_{\rm ta}}{k_B}\right)
    \simeq
    \frac{m(K-m)}{K^2}
    \left[
    \log
    \frac{1-m\epsilon}{(K-m)\epsilon}
    \right]^2 .
\end{equation}
This analytical value was compared with the empirical variance measured from trained diffusion models.

\section{Diffusion model on CIFAR-10}
\subsection{Diffusion protocol and model architecture}

We trained a variance-preserving denoising diffusion probabilistic model (DDPM) on CIFAR-10, following the standard DDPM formulation of Ref.~\cite{ho2020denoising}. The forward process used $T=1000$ diffusion steps with a linear noise variance schedule $\beta_t$ ranging from $10^{-4}$ to $2\times 10^{-2}$. This time-dependent noise schedule can be absorbed into a monotone reparameterization of diffusion time to match the Langevin equation in the main text.

The score network was a U-Net operating on $32\times 32$ RGB images. The base channel width was 128, with channel multipliers $(1,2,2,2)$, two residual blocks per resolution level, and self-attention applied at the $16\times 16$ resolution. Timestep conditioning used sinusoidal embeddings followed by a two-layer multilayer perceptron of width $4\times$ the base channel dimension. Each residual block contained Group Normalization, SiLU nonlinearities, two $3\times 3$ convolutions, and a linear projection of the timestep embedding; skip projections were used when the input and output channel dimensions differed. Downsampling was performed with learned stride-2 convolutions, whereas upsampling used nearest-neighbor interpolation followed by a $3\times 3$ convolution. The bottleneck consisted of a residual block, an attention block, and a second residual block. The output head applied Group Normalization, SiLU, and a final $3\times 3$ convolution back to three channels.

\subsection{Dataset and preprocessing}

We used the CIFAR-10 dataset for training and post-training analyses. Images were converted to tensors and linearly rescaled to the range $[-1,1]$ using channel-wise normalization with mean $(0.5,0.5,0.5)$ and standard deviation $(0.5,0.5,0.5)$. Training data were loaded with a batch size of $128$ and four data-loading workers, with incomplete final batches dropped.

\subsection{Optimization}

Models were trained for $500$ epochs with Adam using a learning rate of $2\times 10^{-4}$. Gradients were clipped to a global norm of $1.0$ before each optimization step. An exponential moving average of the model parameters with decay $0.9999$ was maintained throughout training. Samples were generated from the EMA model.

\subsection{Post-training analyses}

In computing the TAEP for a trajectory, we substituted the finally trained model as the optimum, and estimated the divergence term using Hutchinson's method~\cite{hutchinson1989stochastic}. We randomly sampled $1000$ trajectories for each model to estimate the variance of TAEP. The coverage metric is defined in Ref.~\cite{naeem2020reliable}. To study the geometry of the loss landscape, we computed Hessian matrices of the training loss with respect to selected subsets of parameters. Hessians for the selected parameter vector were accumulated over mini-batches. In separate calculations, we estimated the covariance of SGD noise by evaluating the gradient of the mini-batch loss with respect to the same parameter subsets and forming the centered empirical covariance matrix of these mini-batch gradients.

\subsection{Structured parameter selection}

We restricted the Hessian and SGD covariance analyses to structured subsets of $\sim 1000$ parameters, for which exact dense Hessians remained computationally feasible. Parameter subsets were chosen in a manner that respected the internal organization of the network. For linear layers, we selected subsets of input and output nodes and retained the corresponding connections and biases. For convolutional layers, we selected subsets of input and output channels while retaining the full spatial kernel for each chosen channel pair. For attention blocks, we selected matched channel subsets across the query, key, and value projections. Group-normalization parameters were also selected.

\section{Computational resources}

Training and post-training analyses were performed in PyTorch on NVIDIA A100 GPUs.

\section*{Supplementary information}
The supplementary information contains derivations and figures.

\section*{Code availability}
Code that can be used to reproduce the main text figures is available via GitHub at https://github.com/Max-Snow/stochastic-thermodynamics-of-score-matching-objective-for-diffusion-models.

\section*{Acknowledgements}
We thank Surya Ganguli, Bert (HJ) Kappen, and Haewon Jeong for helpful discussions and comments, which helped improve this paper. The Flatiron Institute is a division of the Simons Foundation.

\bibliography{apssamp}

\backmatter

%\begin{appendices}

%\section{Section title of first appendix}\label{secA1}




%\end{appendices}


%TC:endignore
\end{document}

\preprint{APS/123-QED}

\title{Supplementary Material for the paper "Stochastic Thermodynamics of Score Matching in Diffusion Models"}% Force line breaks with \\

\author{Xuehao Ding}
 \affiliation{Flatiron Institute, Simons Foundation}%Lines 
 %\email{xding@flatironinstitute.org}

 \author{H. T. Quan}
%\email{htquan@pku.edu.cn}
 \affiliation{School of Physics, Peking University}%Lines 
\affiliation{Collaborative Innovation Center of Quantum Matter, Peking University}
\affiliation{Frontiers Science Center for Nano-optoelectronics, Peking University}

\author{Yuhai Tu}
 \affiliation{Flatiron Institute, Simons Foundation}%Lines 
 %\email{ytu@flatironinstitute.org}

\maketitle

\onecolumngrid
\renewcommand{\theequation}{S\arabic{equation}}
\setcounter{equation}{0}

\section{Fluctuating time-asymmetry entropy production}

We discretize Eq. \eqref{forward process} using Stratonovich's convention~\cite{lau2007state}.
\begin{equation}
    x_{n+1}-x_n=\mu F(m)\Delta t+\sqrt{2\mu k_BT\Delta t}\cdot z_n,
\end{equation}
where $m\coloneq\frac{x_{n+1}+x_n}{2}$, $z_n\sim \mathcal{N}(0,I)$. Denote
\begin{equation}
\Delta x_n=x_{n+1}-x_n,
\end{equation}
\begin{equation}
    h \coloneq \Delta x_{n} - \mu F(m)\Delta t,
\end{equation}
we have
\begin{equation}
p(h|x_n)=\frac{1}{\sqrt{2\pi\cdot 2\mu k_BT\Delta t}}\exp\left(-\frac{||h||^2}{2\cdot 2\mu k_BT\Delta t}\right).
\end{equation}
It follows that the forward transition probability is given by~\cite{lau2007state}
\begin{equation}
\begin{aligned}
p(x_{n+1}|x_n)=&p(h|x_n)\det\left(\frac{\partial h}{\partial x_{n+1}}\right)\\
=&p(h|x_n)\det(I-\frac{1}{2}\mu \nabla F(m) \Delta t)\\
=&p(h|x_n)[1-\frac{1}{2}\mu \nabla\cdot F(m) \Delta t+o(\Delta t)],
\end{aligned}
\end{equation}
where the last line uses Jacobi's formula and Taylor expansion.

Similarly, we can calculate the transition probability of the reverse process from $x_{n+1}$ to $x_n$ at time $\widetilde{t}=\tau - t$.
\begin{equation}
\begin{aligned}
\widetilde{p}_\theta(x_n|x_{n+1};\tau-t)=&\frac{1}{\sqrt{2\pi\cdot 2\mu k_BT\Delta t}}\exp\left(-\frac{||-\Delta x_n+\mu F(m)\Delta t-2\mu k_B Ts_\theta(m,\tau-t)\Delta t||^2}{2\cdot 2\mu k_BT\Delta t}\right)\\
& [1+\frac{1}{2}\mu \nabla\cdot F(m) \Delta t-\mu k_BT \nabla\cdot s_\theta(m,\tau-t) \Delta t +o(\Delta t)].
\end{aligned}
\end{equation}
Applying Taylor expansion to $\log(1+x)$ for $x\ll 1$, we have the following.
\begin{equation}
\begin{aligned}
\log \frac{p(x_{n+1}|x_n)}{\widetilde{p}_\theta(x_n|x_{n+1};\tau-t)}=&-\frac{||\Delta x_{n} - \mu F(m)\Delta t||^2}{4\mu k_BT\Delta t}+\frac{||-\Delta x_n+\mu F(m)\Delta t-2\mu k_B Ts_\theta(m,\tau-t)\Delta t||^2}{4\mu k_BT\Delta t}\\
&+\mu k_BT\nabla \cdot s_\theta(m,\tau-t)\Delta t -\mu\nabla\cdot F(m)\Delta t +o(\Delta t)\\
=&s_\theta(m,\tau-t)^\top[\Delta x_n-\mu F(m)\Delta t+\mu k_B Ts_\theta(m,\tau-t)\Delta t]\\
&+\mu k_BT\nabla \cdot s_\theta(m,\tau-t)\Delta t -\mu\nabla\cdot F(m)\Delta t +o(\Delta t).
\end{aligned}
\end{equation}
By integrating along the trajectory, we obtain an expression of the TAEP for a trajectory $x([t])$:
\begin{equation}
\label{time-asymmetry EP expansion}
\begin{aligned}
\Delta s_{ta}=&\Delta s_{sys}+k_B\log\frac{p[x([t])|x(0)]}{\widetilde{p}_\theta[\bar{x}([t])|x(\tau)]}\\
=&\Delta s_{sys}+k_B\int_{0}^{\tau} s_\theta(x,\tau-t)^\top\circ dx\\
&+ \mu k_B\int_{0}^\tau [- F(x)^\top s_\theta(x,\tau-t)+ k_B T||s_\theta(x,\tau-t)||^2 +k_BT\nabla \cdot s_\theta(x,\tau-t)-\nabla\cdot F(x)]dt\\
=& \Delta s_{sys}+k_B\int_{0}^{\tau} s_\theta(x,\tau-t)^\top\sqrt{2\mu k_B T}\circ dW_t\\
&+ \mu k_B^2T\int_{0}^\tau [||s_\theta(x,\tau-t)||^2+ \nabla \cdot s_\theta(x,\tau-t)]dt-\mu k_B\int_0^\tau \nabla\cdot F(x)dt\\
=& \Delta s_{sys}+k_B\int_{0}^{\tau} s_\theta(x,\tau-t)^\top\sqrt{2\mu k_B T} dW_t\\
&+ \mu k_B^2T\int_{0}^\tau [||s_\theta(x,\tau-t)||^2+ 2\nabla \cdot s_\theta(x,\tau-t)]dt-\mu k_B\int_0^\tau \nabla\cdot F(x)dt,\\
\end{aligned}
\end{equation}
where $\Delta s_{sys}$ is defined as Eq. \eqref{system entropy}, the third equality uses Eq. \eqref{forward process}, and the fourth equality converts the Stratonovich integral to the Itô integral.

We now expand the system entropy change as integrals.
\begin{equation}
\label{system EP expansion}
\begin{aligned}
\Delta s_{sys}
=&-k_B\Delta \log p(x,t)\\
=& -k_B\int_0^\tau[\nabla\log p(x,t)]^\top\circ dx - k_B\int_0^\tau\frac{\partial\log p(x,t)}{\partial t} dt\\
=&-k_B\int_0^\tau\mu[\nabla\log p(x,t)]^\top F(x)dt-k_B\int_0^\tau\sqrt{2\mu k_B T}[\nabla\log p(x,t)]^\top dW_t\\
&-\mu k_B^2T\int_0^\tau \nabla^2 \log p(x,t)dt-k_B\int_0^\tau\frac{-\nabla\cdot [\mu F(x)p(x,t)]+\mu k_BT\nabla^2 p(x,t)}{p(x,t)}dt\\
=&\mu k_B\int_0^\tau \nabla \cdot F(x) dt-k_B\int_0^\tau\sqrt{2\mu k_B T}[\nabla\log p(x,t)]^\top dW_t\\
&-\mu k_B^2 T\int_0^\tau [2\nabla^2 \log p(x,t)+||\nabla \log p(x,t)||^2]dt,
\end{aligned}
\end{equation}
where the third equality converts the Stratonovich integral to the Itô integral and uses the Fokker-Planck equation. Combining Eqs. \eqref{time-asymmetry EP expansion}\eqref{system EP expansion}, we finally obtain the expression of $\Delta s_{ta}$.
\begin{equation}
\begin{aligned}
&\Delta s_{ta}[x([t])]\\
=&k_B\sqrt{2\mu k_B T}\int_{0}^{\tau} [s_\theta(x,\tau-t)-\nabla \log p(x,t)]^\top dW_t\\
&+\mu k_B^2T\int_{0}^\tau [||s_\theta(x,\tau-t)||^2+ 2\nabla \cdot s_\theta(x,\tau-t)]dt-\mu k_B^2 T\int_0^\tau [||\nabla \log p(x,t)||^2+2\nabla^2 \log p(x,t)]dt.
\end{aligned}
\end{equation}

\section{Ensemble-averaged time-asymmetry entropy production}

The last term of Eq. \eqref{TAEP expression} is a martingale. Thus, the ensemble-averaged TAEP is given by
\begin{equation}
\label{ensemble nr EP calculation}
\begin{aligned}
&\langle\Delta s_{ta}[x([t])]\rangle\\
=&\mu k_B^2T\int_{0}^\tau\int p(x,t) [||s_\theta(x,\tau-t)||^2+ 2\nabla \cdot s_\theta(x,\tau-t)]dxdt\\
&-\mu k_B^2 T\int_0^\tau\int p(x,t) [||\nabla \log p(x,t)||^2+2\nabla^2 \log p(x,t)]dxdt.
\end{aligned}
\end{equation}
Using integration by parts and assuming that the boundary term is zero, we have
\begin{equation}
\label{integration by parts 1}
\begin{aligned}
\int p(x,t)\nabla \cdot s_\theta(x,\tau-t)dx=-\int p(x,t)\nabla \log p(x,t)\cdot s_\theta(x,\tau-t)dx,
\end{aligned}
\end{equation}
\begin{equation}
\label{integration by parts 2}
\begin{aligned}
\int p(x,t) [||\nabla \log p(x,t)||^2+2\nabla^2 \log p(x,t)]dx=-\int p(x,t) ||\nabla \log p(x,t)||^2 dx.
\end{aligned}
\end{equation}
Substituting Eqs. \eqref{integration by parts 1}\eqref{integration by parts 2} into Eq. \eqref{ensemble nr EP calculation}, we finally obtain
\begin{equation}
\begin{aligned}
\langle\Delta s_{ta}[x([t])]\rangle=&\mu k_B^2T\int_{0}^\tau\int p(x,t) ||s_\theta(x,\tau-t)-\nabla \log p(x,t)||^2dxdt\\
=& \mu k_B^2T\mathcal{L}_{SM}(\theta).
\end{aligned}
\end{equation}

\section{Scenarios where the exact-score approximation is valid}
In the following, we discuss several scenarios where the exact-score approximation, Eq. \eqref{exact score field}, is valid.
\begin{enumerate}
\item Transfer learning, which is a machine learning paradigm that leverages knowledge acquired from a pre-trained source domain to enhance the learning efficiency and predictive performance of a model on a related but distinct task. Suppose that there is a dataset $\mathcal{P}$ with distribution $p(x,0)$ and a related dataset $\mathcal{Q}$ with distribution $q(x,0)$. Suppose that a diffusion model has been fully trained on $\mathcal{Q}$, i.e., Eq. \eqref{exact score field} is satisfied. Now we want to train another diffusion model to learn the distribution of $\mathcal{P}$. Instead of training from scratch, we start from the existing network to transfer knowledge about $\mathcal{Q}$ to the new diffusion model.
\item Generalization. Similarly to transfer learning, the model has been fully trained on the train set $\mathcal{Q}$, and now we test the model on the test set $\mathcal{P}$.
\item Near-optimality. When $\theta$ is in the neighborhood of the optimum $\theta^*$, $s_\theta(x,\tau-t)$ can be well approximated as its projection onto the score-field space~\cite{lai2023fp}, which we denote by $\nabla \log q(x,t)$.
\end{enumerate}

Notice that the evolution of $q(x,t)$ is governed by the same Fokker-Planck equation:
\begin{equation}
    \frac{\partial q(x,t)}{\partial t}=-\nabla[\mu F(x)q(x,t)-\mu k_B T\nabla q(x,t)].
\end{equation}
We substitute $s_\theta(x,\tau-t)=\nabla \log q(x,t)$ into Eq. \eqref{time-asymmetry EP expansion}.
\begin{equation}
\begin{aligned}
\Delta s_{ta}=&\Delta s_{sys}+k_B\int_{0}^{\tau} s_\theta(x,\tau-t)^\top\circ dx\\
&+ \mu k_B\int_{0}^\tau [- F(x)^\top s_\theta(x,\tau-t)+ k_B T||s_\theta(x,\tau-t)||^2 +k_BT\nabla \cdot s_\theta(x,\tau-t)-\nabla\cdot F(x)]dt\\
=& \Delta s_{sys}+k_B\int_{0}^{\tau} [\nabla \log q(x,t)]^\top\circ dx\\
&+ \mu k_B\int_{0}^\tau [- F(x)^\top \nabla \log q(x,t)+ k_B T||\nabla \log q(x,t)||^2 +k_BT\nabla^2 \log q(x,t)-\nabla\cdot F(x)]dt\\
=& \Delta s_{sys}+k_B\int_{0}^{\tau} [\nabla \log q(x,t)]^\top\circ dx\\
&- k_B\int_{0}^\tau \frac{1}{q(x,t)}\nabla[\mu F(x)q(x,t)-\mu k_B T\nabla q(x,t)]dt\\
=&\Delta s_{sys}+k_B\int_{0}^{\tau} \partial_\mu \log q(x,t)\circ dx^\mu,
\end{aligned}
\end{equation}
where the last equality uses the Fokker-Planck equation. Using the second equality of Eq. \eqref{system EP expansion}, we can further express the fluctuating TAEP as
\begin{equation}
\Delta s_{ta}=k_B\int_{0}^{\tau} \partial_\mu \log \frac{q(x,t)}{p(x,t)}\circ dx^\mu.
\end{equation}
Then it follows that the average TAEP rate is given by
\begin{equation}
\label{nr EP rate for transfer learning_si}
\frac{d S_{ta}}{dt}=k_B\int \mathcal{J}^\mu(x,t)\partial_\mu \log \frac{q(x,t)}{p(x,t)}dx,
\end{equation}
where $\mathcal{J}\coloneq (p,J)$.

Using integration by parts and assuming that the boundary term vanishes, Eq. \eqref{nr EP rate for transfer learning_si} can be alternatively written as the derivative of a KL divergence:
\begin{equation}
    \frac{d S_{ta}}{dt}=-k_B\frac{d}{dt}D_{KL}[p(x,t)||q(x,t)],
\end{equation}
which is known to be non-negative~\cite{csiszar1967information}. Intuitively, $p(x,t)$ and $q(x,t)$ both converge to the stationary distribution in the forward process, and thus their KL divergence decreases.

\section{Relation between mean and variance constrained by fluctuation theorem}
\begin{theorem}
Let $(X_n)_{n\ge 1}$ be a sequence of real-valued random variables satisfying the integral fluctuation theorem
\[
\mathbb E\!\left[e^{-X_n}\right]=1
\]
for every $n$. Suppose also that
\[
\mathbb E[X_n]\longrightarrow 0.
\]
Assume moreover that the family $\{X_n^2\}_{n\ge 1}$ is uniformly integrable, i.e.
\[
\lim_{K\to\infty}\sup_n 
\mathbb E\!\left[X_n^2\mathbf 1_{\{|X_n|>K\}}\right]=0.
\]
Then
\[
\operatorname{Var}(X_n)\longrightarrow 0.
\]
\end{theorem}

\begin{proof}
Define
\[
g(x)=x+e^{-x}-1.
\]
By the elementary inequality
\[
e^{-x}\ge 1-x,
\]
we have
\[
g(x)\ge 0
\]
for all $x\in\mathbb R$. Moreover, equality holds if and only if $x=0$.

Using the fluctuation theorem condition,
\[
\mathbb E[e^{-X_n}]=1,
\]
we get
\[
\mathbb E[g(X_n)]
=
\mathbb E[X_n]+\mathbb E[e^{-X_n}]-1
=
\mathbb E[X_n].
\]
Therefore,
\[
\mathbb E[g(X_n)]\longrightarrow 0.
\]

We first show that $X_n\to 0$ in probability. Fix $\varepsilon>0$. Since $g$ is continuous, strictly positive on the set
\[
\{x: |x|\ge \varepsilon\},
\]
and satisfies $g(x)\to\infty$ as $|x|\to\infty$, the number
\[
c_\varepsilon
:=
\inf_{|x|\ge \varepsilon} g(x)
\]
is strictly positive. Hence
\[
g(X_n)\ge c_\varepsilon \mathbf 1_{\{|X_n|\ge \varepsilon\}}.
\]
Taking expectations gives
\[
c_\varepsilon \mathbb P(|X_n|\ge \varepsilon)
\le
\mathbb E[g(X_n)].
\]
Since $\mathbb E[g(X_n)]\to 0$, it follows that
\[
\mathbb P(|X_n|\ge \varepsilon)\to 0.
\]
Thus
\[
X_n\to 0
\]
in probability.

We now upgrade convergence in probability to convergence in $L^2$. By uniform integrability of $\{X_n^2\}$, for every $\eta>0$ there exists $K>0$ such that
\[
\sup_n \mathbb E\!\left[X_n^2\mathbf 1_{\{|X_n|>K\}}\right]<\eta.
\]
Fix also $\delta>0$. Decompose
\[
\mathbb E[X_n^2]
=
\mathbb E\!\left[X_n^2\mathbf 1_{\{|X_n|\le \delta\}}\right]
+
\mathbb E\!\left[X_n^2\mathbf 1_{\{\delta<|X_n|\le K\}}\right]
+
\mathbb E\!\left[X_n^2\mathbf 1_{\{|X_n|>K\}}\right].
\]
The three terms are bounded as follows:
\[
\mathbb E\!\left[X_n^2\mathbf 1_{\{|X_n|\le \delta\}}\right]\le \delta^2,
\]
\[
\mathbb E\!\left[X_n^2\mathbf 1_{\{\delta<|X_n|\le K\}}\right]
\le
K^2\mathbb P(|X_n|>\delta),
\]
and
\[
\mathbb E\!\left[X_n^2\mathbf 1_{\{|X_n|>K\}}\right]<\eta.
\]
Therefore
\[
\mathbb E[X_n^2]
\le
\delta^2
+
K^2\mathbb P(|X_n|>\delta)
+
\eta.
\]
Taking $\limsup_{n\to\infty}$ and using $X_n\to0$ in probability, we obtain
\[
\limsup_{n\to\infty}\mathbb E[X_n^2]
\le
\delta^2+\eta.
\]
Since $\delta>0$ and $\eta>0$ are arbitrary,
\[
\mathbb E[X_n^2]\to0.
\]

Finally,
\[
\operatorname{Var}(X_n)
=
\mathbb E[X_n^2]-(\mathbb E[X_n])^2.
\]
We have shown that $\mathbb E[X_n^2]\to0$, and by assumption $\mathbb E[X_n]\to0$. Hence
\[
\operatorname{Var}(X_n)\to0.
\]
\end{proof}

\section{Variance of per-trajectory and per-sample loss}

In practice, the score-matching objective, Eq. \eqref{score matching objective}, is usually decomposed into the per-sample loss as follows.
\begin{equation}
\begin{aligned}
    \mathcal{L}_{SM}(\theta)&=\mathbb{E}_{x_0\sim p(x,0)}[\mathcal{L}_{DSM}(\theta,x_0)]+const.\\
     \mathcal{L}_{DSM}(\theta,x_0)&\coloneq \int_0^\tau\langle||s_\theta(x,\tau-t)-\nabla \log p(x,t|x_0)||^2\rangle_{x\sim p(x,t|x_0)} dt,
\end{aligned}
\end{equation}
which is known as the denoising score-matching objective~\cite{ho2020denoising}. In particular, when $F(x)$ is a linear function in $x$, as in variance-exploding and variance-preserving settings, the conditional score $\nabla \log p(x,t|x_0)$ is proportional to the corrupting noise added to the sample. People usually use the Monte Carlo method to estimate the per-sample loss and minimize it using optimizers.

\begin{figure}[ht]
  \centering
  \includegraphics[width=0.4\textwidth]{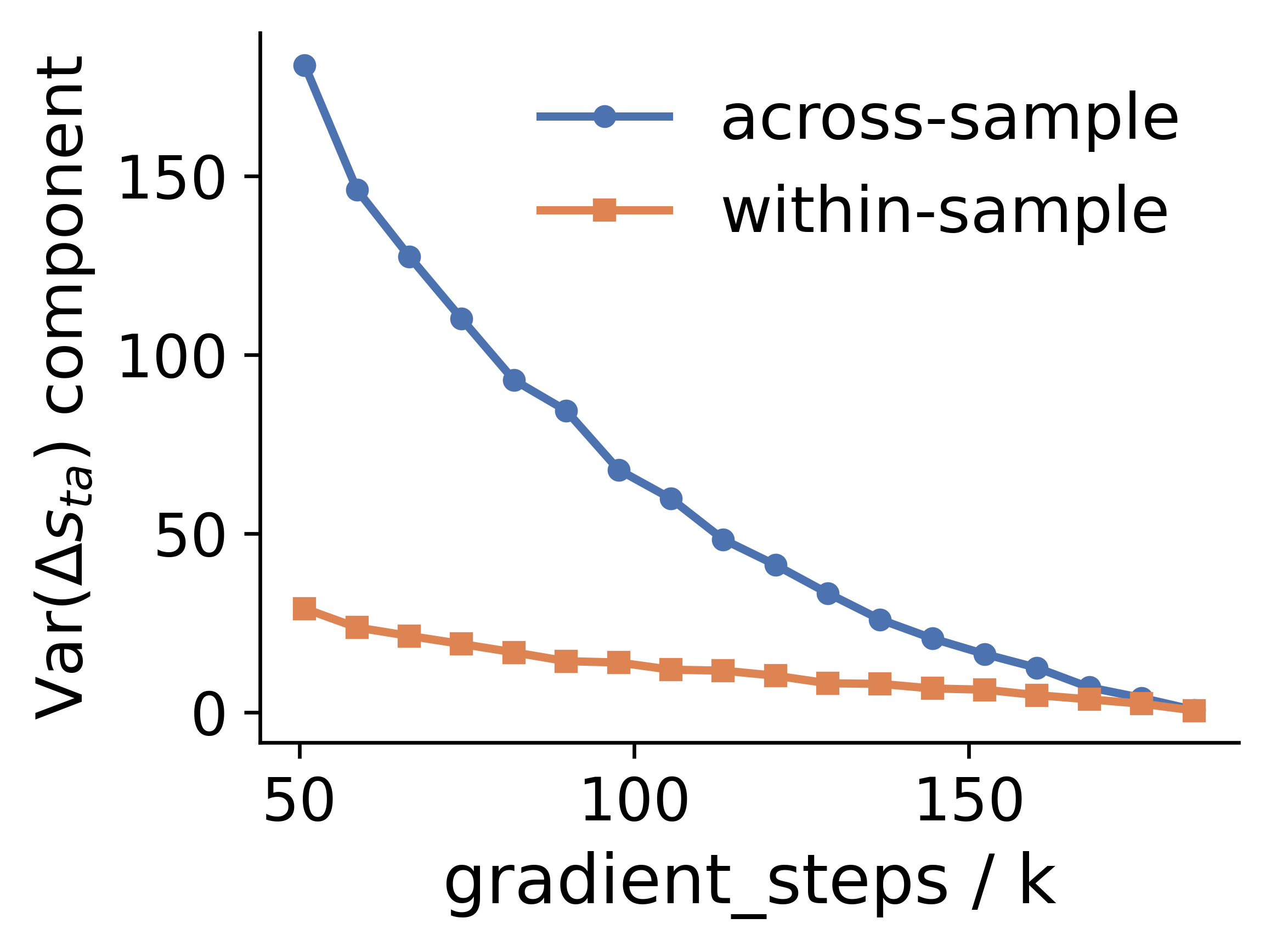}
  \caption{Across-sample variance and within-sample variance versus the gradient step, which correspond to the two terms in Eq. \eqref{variance decomposition}.}
  \label{var component}
\end{figure}

We note that, up to constants, the per-sample loss is proportional to the average of the trajectory TAEP conditioned on the sample:
\begin{equation}
\begin{aligned}
    &\langle \Delta s_{ta}[x([t])]|x(0)=x_0\rangle\\
    =&\mu k_B^2T\int_{0}^\tau\int p(x,t|x_0) [||s_\theta(x,\tau-t)||^2+ 2\nabla \cdot s_\theta(x,\tau-t)]dxdt+const.\\
    =&\mu k_B^2T\int_{0}^\tau\int p(x,t|x_0) [||s_\theta(x,\tau-t)||^2- 2\nabla\log p(x,t|x_0) \cdot s_\theta(x,\tau-t)]dxdt+const.\\
    =&\mu k_B^2T\int_{0}^\tau\int p(x,t|x_0) ||s_\theta(x,\tau-t)-\nabla\log p(x,t|x_0)||^2dxdt+const.\\
    =& \mu k_B^2T \mathcal{L}_{DSM}(\theta, x_0)+const.
\end{aligned}
\end{equation}
Treating the trajectories starting from the same sample as a group, the variance of the TAEP can be decomposed into the variance of group means and the within-group variance:
\begin{equation}
\label{variance decomposition}
\text{Var}(\Delta s_{ta})=(\mu k_B^2T)^2\text{Var}[\mathcal{L}_{DSM}(\theta, x_0)]+\mathbb{E}[\text{Var}(\Delta s_{ta}|x(0)=x_0)].
\end{equation}
We numerically compare the magnitudes of these two components throughout the training process (Fig. \ref{var component}), indicating that the per-sample loss variance is the dominant component of the per-trajectory loss variance.

%TC:endignore'
%\bibstyle{../sn-mathphys-num}
\bibliography{apssamp}% Produces the bibliography via BibTeX.